\documentclass[preprint,showpacs,preprintnumbers,amsmath,amssymb,32]{revtex4}

\usepackage{multido}
\usepackage[active]{srcltx}
\usepackage{graphicx}
\usepackage{dcolumn}
\usepackage{bm}
\usepackage{setspace}

\begin{document}

\title{Surface roughness effect on ultracold neutron interaction with a wall, 
and implications for computer simulations}

\author{A.~Steyerl*, S.~S.~Malik, A.~M.~Desai, C. Kaufman}
\address{Department of Physics, University of Rhode Island, Kingston, RI 02881, U. S. A.}

\begin{abstract}
 We review the diffuse scattering and the loss coefficient
in ultracold neutron reflection from slightly rough surfaces, report
a surprising reduction in loss coefficient due to roughness, and
discuss the possibility of transition from quantum treatment to ray
optics. The results are used in a computer simulation of neutron storage in a
recent
 neutron lifetime experiment that reported a large discrepancy of neutron
 lifetime
 with the current particle data value. Our partial re-analysis suggests the 
possibility of systematic effects that were not included in this publication.
 \\
 
\end{abstract}

\email{asteyerl@mail.uri.edu}

\maketitle
\section{\label{sec:level1} Introduction}

The scattering and absorption of cold and ultracold neutrons (UCN)
at slightly rough surfaces resembles the roughness problem in light
and X-ray optics [1]. Surface roughness induces losses in neutron
guide tubes and affects the behavior of a UCN gas in a trap. We
investigate especially the relation between a realistic account of
roughness in computer simulations of UCN storage and a reliable data
interpretation with respect to the neutron lifetime. The issue has
gained added importance since a new UCN storage based lifetime value
six standard deviation away from the world average value [2] was
published in 2005 [3].

Most theoretical analyses of diffuse cold and ultracold neutron scattering at
surfaces with small roughness (e.g., [4-7]) are based on first-order
perturbation theory. A possible extension to macroscopic ray optics was
presented in [4]. Carrying the analysis to second order perturbation theory,
Ignatovich [7] derived the roughness effect on the loss coefficient for
reflection at a slightly absorbing wall material. The analysis was complex, and
the result has, apparently, not been numerically evaluated and applied so
far. We used a more direct approach, verified and quantified the result of [7],
and in the process obtained a derivation of the "Debye-Waller factor"
(describing the attenuation of the specular beam due to roughness), that is
consistent with standard expressions but requires fewer assumptions about the
roughness characteristics. This appears important since in practical cases very
little is known about the roughness parameters of a given surface. Finally,
 we address the question to what extent a transition to a macroscopic picture is possible. 

\section{\label{sec:level2} Correlation functions describing a rough surface}

Small irregular deviations of a slightly rough surface from the
plane geometry are usually described by a height-height correlation
\begin{equation}
f(\boldsymbol\delta)=\lim_{A \to \infty}\frac{1}{\textit{A}}
\int_{A}
 \xi(\boldsymbol\rho)\xi(\boldsymbol\rho+\boldsymbol\delta)\,d^2\boldsymbol\delta
\end{equation}
where $\xi(\boldsymbol\rho) $ is the random elevation at point
$\boldsymbol\rho=(\textit{x},\textit{y})$ of the plane surface above
its average $\textit{z}=0$. \textit{A} is the illuminated surface
area and $\boldsymbol\delta$ is the displacement vector between two
points. Among common models [4-6] we will concentrate on a Gaussian
correlation for solid surfaces with mean-square roughness
$\textit{b}^2= \left<\xi^{2}\right>$:
\begin{equation}
f_G(\boldsymbol\delta)= f_G(\delta)=b^2 \exp[-\delta^2/(2w^2)]
\end{equation}
and a `$K_{0}$-model' [5] for liquids and possibly glasses retaining
characteristics of a liquid below the glass transition:
\begin{equation}
f_K(\boldsymbol\delta)= f_K(\delta)=\textit{b}^{2} \frac
{K_{0}\{[(\delta^{2}+\delta^{2}_{0})/(2\textit{w}^{2})]^{1/2}\}}{K_{0}[\delta_{0}/(\textit{w}\sqrt{2})]}
\end{equation}

The latter form contains the modified Bessel function $K_{0}$ and
has been proposed in [6] as a modification of a liquid model with
logarithmic short-range divergence [5] (which would imply a
divergence of $\textit{g}(\delta)$, see below) to account for
smoothing due to surface tension. Smoothing is achieved by applying
a short-range cutoff, $\delta_{0}$, in addition to the long-range
cutoff \textit{w} that  is used for both models. In [4, 6] we have
emphasized the importance of the slope-slope correlation for the
surface gradient $\boldsymbol\chi=\nabla\xi(\boldsymbol\rho):$
\begin{equation}
\textit{g}(\boldsymbol\delta)=\textit{g}(\delta) =\lim_{A \to
\infty}\frac{1}{A} \int_{A}
 \boldsymbol\chi(\boldsymbol\rho)\cdot\boldsymbol\chi(\boldsymbol\rho+\boldsymbol\delta)\,d^2\boldsymbol\delta=-\nabla^{2}\textit{f}(\delta)
\end{equation}
which is determined by $f(\delta)$ through the Laplace operator
$\nabla^{2}$. This can be verified using Gauss's divergence
theorem. For the Gaussian model:
\begin{equation}
\textit{g}_{G}(\delta)=\alpha_{G}^2\left(1-
\frac{\delta^{2}}{2\textit{w}^{2}}\right)\exp[-\delta^2/(2w^2)]
\end{equation}
 with mean-square slope $\alpha_{G}^{2}=\left<\chi^{2}\right>= \textit{g}(0)
= 2\textit{b}^{2}/\textit{w}^{2}$. For the `$K_{0}$-model' [6]
\begin{equation}
\textit{g}_{K}(\delta)=\alpha_{K}^{2}\frac {P\{
[(\delta^{2}+\delta^{2}_{0})/(2\textit{w}^{2})]^{1/2}\}}{P
[\delta_{0}/(\textit{w}\sqrt{2})]}
\end{equation}
with
\begin{equation*}
P(\nu)=\frac{\delta_{0}^2/(2w^2)}{\nu^2}K_{2}(\nu)-K_{0}(\nu)
\end{equation*}
and
\begin{equation*}
\alpha_{K}^{2}=\frac{\textit{b}^2}{\textit{w}^2}\frac{K_{1}(t)}{tK_{0}(t)}
\end{equation*}

To facilitate comparison with the Gaussian model we choose
$\delta_{0}$ such that for given \textit{b} and \textit{w} the
mean-square slopes become identical:
$\alpha_{G}^{2}=\alpha_{K}^{2}=\alpha^{2}$. This requires
$\textit{t}=\delta_{0}/(\textit{w}\sqrt{2})=0.7709$ which is the
solution of $K_{2}(t)/K_{0}(t)=5$.

Fairly smooth surfaces with small $\alpha^{2}$ are best
characterized by $g(\delta)$. For surfaces with tips and sharp edges
the curvature-curvature correlation function
\begin{equation}
h(\boldsymbol\delta)=h(\delta)= \lim_{A \to
\infty}\frac{1}{A}\int_{A}
 \kappa(\boldsymbol\rho)\kappa(\boldsymbol\rho+\boldsymbol\delta)\,d^2\boldsymbol\delta=
 \frac{1}{4}\nabla^{2}\nabla^{2}f(\boldsymbol\delta)
\end{equation} \\
may not be negligible, either. For $\alpha^{2}<<1$ the mean surface
curvature is given by
$\kappa=\frac{1}{2}\left(\frac{\partial^{2}\xi}{\partial
x^{2}}+\frac{\partial^{2}\xi}{\partial y^{2}}\right)=\frac{1}{2}
\nabla^{2}\xi$ and represents the mean curvature in any two
orthogonal in-plane directions \textit{x} and \textit{y}. The
operations in (7) can be performed using the properties of Bessel
functions. The results are given in Appendix A. The mean values are
\begin{equation*}
\kappa_{G}^2=h_G(0)=2\textit{b}^2/\textit{w}^4=\alpha^2/\textit{w}^2
\end{equation*}
and
\begin{equation*} \kappa_{K}^2=h_K(0)=2.103 \alpha^2/\textit{w}^2
\end{equation*}
for $\textit{t}= 0.7709$, as before.

We add a note related to calculations of `Debye-Waller factors'
describing the decrease of specular intensity, for larger momentum
transfer, due to destructive interference within the rough layer.
These are based on probability distributions for height,
$\textit{p}(\xi)$, or slope, $\textit{p}(\eta)$, etc., that require
additional assumptions about surface properties. For instance, the
commonly used Gaussian form
$\textit{p}(\xi)=b^{-1}(2\pi)^{-1/2}\exp[-\xi^{2}/(2b^{2})]$, which
is normalized and has second moment $\left<\xi^{2}\right>= b^{2}$,
is \textit{not} implied by the Gaussian form of $f(\delta)$; only
the second moment is common. While the correlations (4), (7) and
higher are unambiguously determined by $f(\delta)$ alone, all the
higher moments for $\textit{p}(\xi)$ constitute additional
assumptions. For real surfaces even \textit{b} and \textit{w} are
usually not well known. We show below that from $f(\delta)$ alone a
Debye-Waller factor can be determined, but only up to order $b^{2}$,
if the perturbation approach for the rough wall is carried to second
order. Other assumptions usually made to derive 'Debye-Waller
factors' will also be discussed.

We will also point out that asymptotic aspects of roughness-induced
scattering and absorption can be obtained from mean values $(b^{2},
\alpha^{2}, etc.)$ only, independently of the details of the models
used (Gaussian or `$K_{0}$' or similar). Going to second order
perturbation is both necessary and sufficient for a full account of
reflection and absorption up to quadratic terms $(b^{2}$). Higher
perturbations are at least of order $b^{4}$ for symmetrical roughness where 
$f(\boldsymbol\delta)=f(-\boldsymbol\delta)$.
\section{\label{sec:level3} Perturbation approach to UCN interaction
with a rough wall}

As usual [4-6] a wall with micro-roughness may be divided into a
volume $V_{0}$ with an ideally smooth wall at $z=0$, and a thin
roughness volume $V_{1}$ with partly positive and partly negative
thickness $\xi(\boldsymbol\rho)$,
 where $\left<\xi\right>=0$. For
an incident plane wave $\psi_{i}(\textbf{r})=
\exp(i\textbf{k}_{i}\cdot\textbf{r})= \exp(-ikz\cos\theta_{i})
\exp(ikx\sin\theta_{i})$ approaching the surface from the vacuum
side ($\textit{z}>0$) the neutron wave function $\psi(\textbf{r})$
may be split into the unperturbed part
 $\psi_{0}(\textbf{r})$ for the plane surface and a small perturbation
$\psi_{1}(\textbf{r})$:
\begin{equation}
\psi(\textbf{r})=\psi_{0}(\textbf{r})+\psi_{1}(\textbf{r})\cong\psi_{0}(\textbf{r})+\psi^{(1)}_{1}(\textbf{r})+\psi^{(2)}_{1}(\textbf{r})
\end{equation}
The first-order perturbation is obtained [4] as
\begin{equation}
\psi^{(1)}_{1}(\textbf{r})=-\textit{q}_{0}\int
G(\textbf{r}|\textbf{r}')\psi_{0}(\textbf{r}')d^3\textbf{r}'
\end{equation}
where the scattering potential $\textit{q}_{0}=Na(1-i\eta)=
\textit{k}_{c}^{2}/4\pi =(\textit{k}_{c0}^{2}/4\pi)(1-i\eta)$ is
determined by $\textit{Na}$, a mean value for the number \textit{N}
of atoms times their bound-atom scattering length a. The imaginary
part given by $\eta=-Im(\textit{q}_{0})/Re(\textit{q}_{0})$ takes
into account loss processes (nuclear capture, inelastic and, except
for UCN, incoherent-elastic scattering). $\eta$ is small for wall
materials of interest for UCN storage. $\textit{k}_{c0}$ denotes the
critical wave number for total reflection at normal incidence on a
flat wall of uniform scattering potential.

The Green's function $G(\textbf{r}|\textbf{r}')$ in (9) satisfies
the equation
\begin{equation}
\nabla^2G(\textbf{r}|\textbf{r}')+K^{2}(\textbf{r})G(\textbf{r}|\textbf{r}')=
-4\pi\delta(\textbf{r}-\textbf{r}')
\end{equation}
where \textit{K}(\textbf{r}) is the wave number: $K=\textit{k}$ in
vacuum, and $K=(\textit{k}^{2}-\textit{k}_{c}^{2})^{1/2}$ within the
medium. For negligible refraction,
$G(\textbf{r}|\textbf{r}')=\exp(ik|\textbf{r}-\textbf{r}'|)/|\textbf{r}-\textbf{r}'|$.
For the general expressions see Appendix B. Asymptotic expressions
for $G(\boldsymbol\rho,z|\boldsymbol\rho',z')$ in the limit of large
\textit{r} have been used in [4].

This method can readily be extended to obtain the second and higher
order perturbations. To obtain the $(n+1)^{th}$ order correction to
$\psi_{1}(\textbf{r})$ we insert $\psi_{1}^{(n)}(\textbf{r})$ in the
integral of (9), and below we will apply this method to extend
roughness scattering to second order. Going to third or higher order
would require correlations of higher than the second order.

 Instead of the
'distorted Green's function method' of [4] used here, an
alternative Distorted Wave Born Approximation (DWBA) was used in
[5]. In [6] it was shown that the two methods are equivalent, with
one important caveat: A conventional 'Born approximation' uses the
far-field Green's function expansion to obtain the factor
$\exp[i(\textbf{k}-\textbf{k}_{i})\cdot\textbf{r}^\prime] $ inside the
integral. This gives an adequate description only in first order.
Second order calculations require also the near field of the Green's
function, at $\textit{z} \approx 0$, in the integral of (9) to
obtain $\psi_{1}^{(1)}(\textbf{r})$ at the rough surface. A
variation of this method was used earlier in [7].

\section{\label{sec:level4}Geometry, basics and main results}

For a wave incident with a wave number \textit{k} in the
(\textit{zx}) plane perpendicular to the wall at polar angle
$\theta_{i}$ (measured from the wall normal), Fig. 1 shows
projections of incident (i), reflected (r) and (elastically)
scattered (s) wave vectors onto the (\textit{xy}) plane. The
scattered beam at solid angle $(\theta,\varphi)$ is characterized by
the in-plane momentum transfer parallel to the wall,
$\textbf{q}=(\textbf{k}_{s}-\textbf{k}_{r})_{par}$, where
\begin{equation}
\textit{q}=k \left[(\sin\theta-\sin\theta_{i})^{2}+ 4\sin\theta
\sin\theta_{i} \sin^{2}\frac{\varphi}{2}\right]^{1/2}
\end{equation}
\begin{figure}
\center
\includegraphics[clip=true,width=4.0in,angle=0]{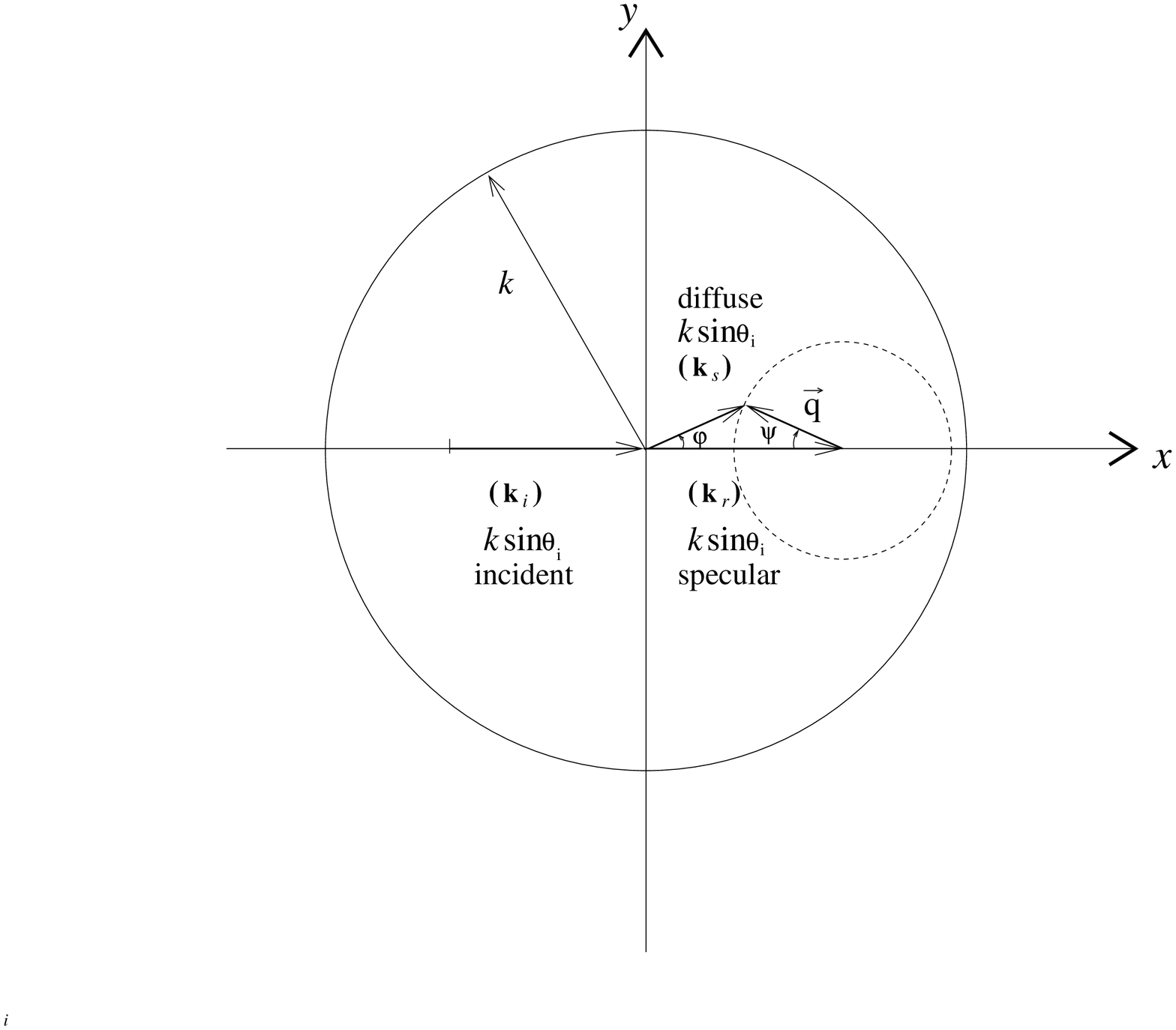}
\caption{For elastic wall interaction of UCN with wave number
\textit{k} we show the projections onto the surface plane of
incident, ($\textbf{k}_{i}$), mirror-reflected, ($\textbf{k}_{r}$),
and diffusely scattered, ($\textbf{k}_{s}$), wave vectors. The
in-plane wave vector transfer \textbf{q} is the component of
$\textbf{k}_{s}$ - $\textbf{k}_{r}$ parallel to the surface.  The
vectors are bracketed to indicate projection onto the
\textit{xy}-plane. For elastic scattering the end points of
\textbf{q} are confined to the circle of radius \textit{k}.}
\end{figure}
Since we consider only elastic scattering, the end point of
\textbf{q} is restricted to the area inside the circle with radius
\textit{k}. For the part of $\psi$ containing the outgoing wave and the evanescent wave inside the medium we
use the expansion
\begin{equation}
\psi_{out}(\textbf{r})=\psi_{0r}(\textbf{r})+
\psi_{1}^{(1)}(\textbf{r})+ \psi_{1}^{(2)}(\textbf{r})
\end{equation}
where
\begin{equation*}
\psi_{0r}(\textbf{r})= \left\{
\begin{array}{rl}
R(\theta_{i}) \exp(i kz\cos\theta_{i})\exp(i
kx\sin\theta_{i});z>0\\S(\theta_{i}) \exp(\kappa_{i}z)\exp(i
kx\sin\theta_{i});z<0
\end{array} \right.
\end{equation*}
is the unperturbed wave field for the flat wall, with
 \begin{equation*}
R(\theta_{i})=\frac
{k\cos\theta_{i}-i\kappa_{i}}{k\cos\theta_{i}+i\kappa_{i}}
 \end{equation*}
and
\begin{equation*}
S(\theta_{i})=1+R(\theta_{i});\hspace{0.2in}
\kappa_{i}=\sqrt{k_{c}^{2}-k^2\cos^{2}\theta_{i}} \,\,\, .
 \end{equation*} 
$\psi_{1}^{(1)}(\textbf{r})$ and $\psi_{1}^{(2)}(\textbf{r})$ denote
the first and second-order perturbation terms.

 Keeping terms $\sim \eta^{1}$ in\textit{ R}, \textit{S}, etc. we obtain the absorption
 corrections to the specular and scattered beams and, by comparison
 of net outgoing to incoming intensities, the loss coefficient for
 the rough wall. For details of the analysis see Appendix C. Here we
 summarize the main results for the intensities normalized to the
 incident flux. Taking the squared magnitude of (12) (outgoing waves
 only) results in cross terms labeled by (mn) where m,n = 0,1,2
 stands for the order of perturbation:
\begin{description}

\item[(a)](0,0): Specular reflectivity for the flat wall, derived from
 $|\psi_{0}|^{2}$: $\textit{I}_{(00)}= |R(\theta_{i})|^{2} = 1-\mu_{0}(\theta_{i})$ where $\mu_{0}(\theta_{i})=2\eta
 \textit{k}\cos(\theta_{i})/\kappa_{i0}$ is the absorption
 coefficient for incidence at angle $\theta_{i}$ on a flat wall [7-9],
 with $\kappa_{i0}=(\textit{k}_{c0}^{2}-\textit{k}^{2}\cos^{2}\theta_{i})^{1/2}$.
   \item[(b)] (0,1): Interference term with first-order perturbation for
 specular reflection, arising from
 $2Re\left(\psi_{0r}^{*}\psi_{1}^{(1)}\right)$:
 \begin{equation}
\textit{I}_{(01)}=-\mu_{0}(\theta_{i})\textit{k}_{c0}^2\textit{b}^2
 \end{equation}
For $\eta=0$, there is no first-order interference. This shows that
a
 'Debye-Waller factor' for specular beam attenuation cannot be
 obtained from first-order perturbation.
   \item[(c)](1,1): The first order intensity scattered into unit solid
 angle at $(\theta,\varphi)$ is given by the outgoing flux with
 density $|\psi_{1}^{(1)}(\textbf{r})|^{2}$, normalized to
 the incident flux, with the result (for UCN)
 \begin{equation}
\textit{I}_{(11)}= 4\textit{k}^4
\cos\theta_{i}\cos^{2}\theta\textit{F}(q)\left[1-\eta\left(\frac{\textit{k}\cos\theta_{i}}{\kappa_{i0}}+\frac{\textit{k}\cos\theta}{\kappa_{0}}\right)\right]
 \end{equation}
where
\begin{equation}
\textit{F}(\textbf{q})=\textit{F}(q)=\frac{1}{(2\pi)^{2}} \int_{A}
\textit{f}(\delta)
e^{-i\textbf{q}\cdot\boldsymbol\delta}d^{2}\boldsymbol\delta
 \end{equation}
is the Fourier transform of the height-height correlation function
 (1) and represents the roughness spectrum [4]. For $\eta=0$, and for
 UCN (rather than cold neutrons in general) Eq. (14) agrees with
 Eq. (20) of [4]. Removing common constants, we write
 $\textit{F(q)}=[\textit{b}^{2}\textit{w}^{2}/(2\pi)]\textit{L}(q)$, where
\begin{subequations}
\begin{align}
 L(q)&= \exp(-q^{2}\textit{w}^{2}/2) \textmd{ for the Gaussian model,
and} \label{second} \\
L(q)&=
\frac{2t}{(1+2q^{2}\textit{w}^{2})^{1/2}}\frac{K_{1}[t(1+2q^{2}\textit{w}^{2})^{1/2}]}{K_{0}(t)}
\textmd{ for the `$K_{0}$-model' with t = 0.7709}. \label{third}
\end{align}
\end{subequations}
 From (14) we obtain the total diffusely scattered
intensity, up to
 terms $\sim \textit{b}^{2}$ and $\eta^{1}$:
 \begin{equation}
p_{D}=p_{D0}-2k^{4}\cos\theta_{i} \int_{(2\pi)} d\Omega
\cos^{2}\theta[\mu_0(\theta_{i})+\mu_0(\theta)]F(q)
 \end{equation}
 where
\begin{equation}
p_{D0}=4k^{4}\cos\theta_{i} \int_{(2\pi)} d\Omega \cos^{2}\theta
F(q)
\end{equation}
is the total scattered intensity for $\eta=0$, and
 $\mu_{0}(\theta)=2\eta \textit{k}\cos\theta/(\textit{k}_{c0}^{2}-\textit{k}^{2}\cos^{2}\theta)^{1/2}$
 is the flat-wall loss coefficient for angle $\theta$. For
 $\alpha<<1$ the scattered intensity $\textit{I}_{(11)}$ forms a halo around the
 specular beam with small width of  order $\alpha$, as expected
 from ray optics.
\item[(d)](0,2): Interference of specular reflection with second-order
 perturbation, arising from $2Re\left(\psi_{0r}^{*}\psi_{1}^{(2)}\right)$:
\begin{equation}
I_{(02)}=-p_{D0}[1-\mu_0(\theta_{i})]+k^{2}k_{c0}^{2}\cos\theta_{i}\int_{(2\pi)}
d\Omega \mu_{0}(\theta) F(q)
\end{equation}
For $\eta=0$, this term is required to satisfy unitarity. Up to
order $\textit{b}^{2}$, the total outgoing intensity (specular plus
diffuse) equals the incoming intensity. In other words, the
'Debye-Waller' attenuation factor DWF for the specular beam equals
$1-\textit{p}_{D0}$, as it should. Carrying out the integration for
$\textit{p}_{D0}$ we find that the DWF is close to but not identical
to the factor
$\exp(-4\textit{b}^{2}\textit{k}^{2}\cos^{2}\theta_{i})\sim
1-4\textit{b}^{2}\textit{k}^{2}\cos^{2}\theta_{i}$  given in [5]. At
this point it should also be mentioned (as is in [5]) that to arrive
at the common Gaussian form for a DWF one has to make the drastic
assumption that the wave inside the medium is given by the same
function as the wave outside the medium, throughout the rough layer.
This is not plausible for larger roughness, especially not for UCN, since only
$\psi$ and its first derivative $\partial\psi/\partial z$ are
continuous at the surface.
   Using (13) and (19) we obtain the Debye-Waller factor for an absorbing, rough
  wall: $\textrm{DWF}=|R|^2+I_{(01)}+I_{(02)}$ 
(up to terms $\sim b^2$ and $\sim \eta$).
   \item[(e)] Combining the terms (0,0), (0,1), (1,1) and (0,2) we obtain the
absorption coefficient for a rough wall:\\
 $\mu(\theta_{i})$ =
(incoming flux - outgoing flux)/(incoming flux) = 1 -
$\left(\textit{I}_{(00)}+\textit{I}_{(01)}+\textit{I}_{(02)}\right)
- \int_{(2\pi)}d\Omega \textit{I}_{(11)}$
\begin{equation}
=\mu_{0}(\theta_{i})(1+k_{c0}^{2}b^{2})-2k^{4}\cos\theta_{i}\int_{(2\pi)}d\Omega\cos^{2}\theta[\mu_0(\theta_{i})-\mu_0(\theta)]F(q)-k^{2}k_{c0}^{2}\cos\theta_{i}\int_{(2\pi)}d\Omega
\mu_{0}(\theta)F(q)
\end{equation}
This agrees with the result derived in [7] using a somewhat
different approach. Numerical results were not given in [7].
Higher-order perturbation would yield only terms $\sim b^n$ with
$n>2$ .
\end{description}

\section{\label{sec:level5}Approximations and Numerical Results}
Numerical evaluation of the double integrals over $\theta$ and
$\varphi$ in (17)-(20) is greatly facilitated if one of the
integrations can be performed analytically. For the Gaussian model,
the $\varphi$-integration leads to the Bessel function $I_{0}$, but
no analytical solution is known for the `$K_{0}$-model'. Using the
transformation
$\textit{k}^{2}d\Omega\cos\theta=d^{2}\textbf{q}=\textit{k}^{2}\nu
d\nu d\psi$, with $\textit{q}=\textit{k}\nu$, the $\psi$-integration
can be performed analytically for any roughness model since
$\textit{F}(\textit{q})$ only depends on $\nu$, not on $\psi$. It is
evident from Fig. 1 that for $\nu<1-\textit{s}_{i}$ (with
$\textit{s}_{i}=\sin\theta_{i})$, the $\psi$-integration runs from
$-\pi$ to $+\pi$. For $1-\textit{s}_{i}<\nu<1+\textit{s}_{i}$, the
limits are $\pm\psi_{u}$ with $ \sin^{2}(\psi_{u}/2))=[1-(\nu
-s_{i})^2]/(4\nu s_{i})$. Expressing $\cos\theta$ in terms of $\nu$
and $\psi$ as $\cos^{2}\theta=1-(s_{i}-\nu)^2-4\nu
s_{i}\sin^{2}(\psi/2)$, the $\psi$-integrations in (17)-(20) can be
performed analytically in terms of elliptic integrals, as shown in
Appendix D. 

Before proceeding with numerical results, we point out the special
case of fairly smooth surfaces with small mean-square slope
$(\alpha^{2}<<1)$ and a UCN wavelength small on the scale of the
lateral correlation length $\textit{w}$, viz.
$\textit{k}\textit{w}>>1$. Smooth surfaces are expected to form when
a special low-temperature oil 'LTF' [10] is sputtered onto a cold
surface, then thermally cycled by slow liquefaction and re-freezing,
as in [3]. Under these circumstances, $\textit{F}(\nu)$ is very
small for $\nu>>(\textit{k}\textit{w})^{-1}$ and the
$\nu$-integration can be extended to $\infty$. In this case,
model-independent results are obtained as follows: The analytical
expressions for the $\psi$-integrals are expanded for small $\nu$ as
$\textit{I}_{\psi} = a_{0}+a_{2}\nu^{2}+a_{4}\nu^{4}+...$ and we use
the identities
\begin{equation}
2\pi \displaystyle\int^\infty_0 qdqF(q)=f(0)=b^{2}
\end{equation}
\begin{equation}
2\pi \displaystyle\int^\infty_0 q^{3}dqF(q)=g(0)=\alpha^{2}
\end{equation}
\begin{equation}
2\pi \displaystyle\int^\infty_0 q^{5}dqF(q)=4h(0)=4\kappa^{2}
\end{equation}
which follow from the properties of Fourier transform (15) and
relations (4) and (7) between
$\textit{f}(\delta),\textit{g}(\delta)$ and $\textit{h}(\delta)$.

In this way asymptotic expressions are obtained for
$\textit{p}_{D0}$ and
$\Delta\mu(\theta_{i})/\mu_{0}(\theta_{i})=[\mu(\theta_{i})-\mu_{0}(\theta_{i})]/\mu_{0}(\theta_{i})$:
\begin{equation}
p_{D0}(\theta_{i})=4k^{2}b^{2}\left\{{c_{i}^{2}-\frac{1+c_{i}^2}{2c_{i}^2}\frac{1}{(k
\textit{w})^{2}}+ O[(kw)^{-4}]}\right\}
\end{equation}
and
\begin{equation}
\Delta\mu(\theta_{i})/\mu_{0}(\theta_{i})=-\frac{\alpha^{2}}{2(1-\zeta^{2}c_{i}^{2})}
\left\{1-\frac{1}{2}\frac{s_{i}^{2}}{c_{i}^2}\frac{2+\zeta^{2}c_{i}^2}{1-\zeta^{2}c_{i}^2}+
O[(kw)^{-2}]\right\}
\end{equation} \\
where $\zeta=\textit{k}/\textit{k}_{c0}\leq 1$ and
$c_{i}=\cos\theta_{i}$. For perturbation theory to be valid,
$\textit{p}_{D0}(\theta_{i})$ must be $<<1$ over the entire range of
$\theta_{i}$.

Taking only the first term in the curly bracket
$(\textit{p}_{D0}\approx
4\textit{b}^{2}\textit{k}^{2}\cos^{2}\theta_{i})$, Eq. (24) is
consistent with the usual DWF with expansion
$(1-4\textit{b}^{2}\textit{k}^{2}\cos^{2}\theta_{i}+...)$, which is
the standard Born approximation result of [5] (see also [12]). Note that for
$\eta=0$ the loss of specular intensity, 1-DWF, equals
$\textit{p}_{D0}(\theta_{i})$ [6], thus particle number is
conserved. Figure 2 compares approximation (24) (first term only)
with the exact numerical result for $\textit{k}_{c0} =0.0802$
nm$^{-1}$, $\alpha=10^{-3}$ and $\textit{w}=2$ $\mu$m ($b=\alpha
\textit{w}/\sqrt{2}=1.4$ nm). The critical value $\textit{k}_{c0}$
is for the LTF oil in its glass state at $\sim 110$ K used in [3].
The agreement is excellent, indicating that for these parameters the
Born approximation for the DWF is adequate. The numerical values of
$\textit{p}_{D0}$ for the Gaussian model are very similar to those
of the $K_{0}$-model used for Fig. 2.

\begin{figure}
\includegraphics[clip=true,width=7in,angle=0]{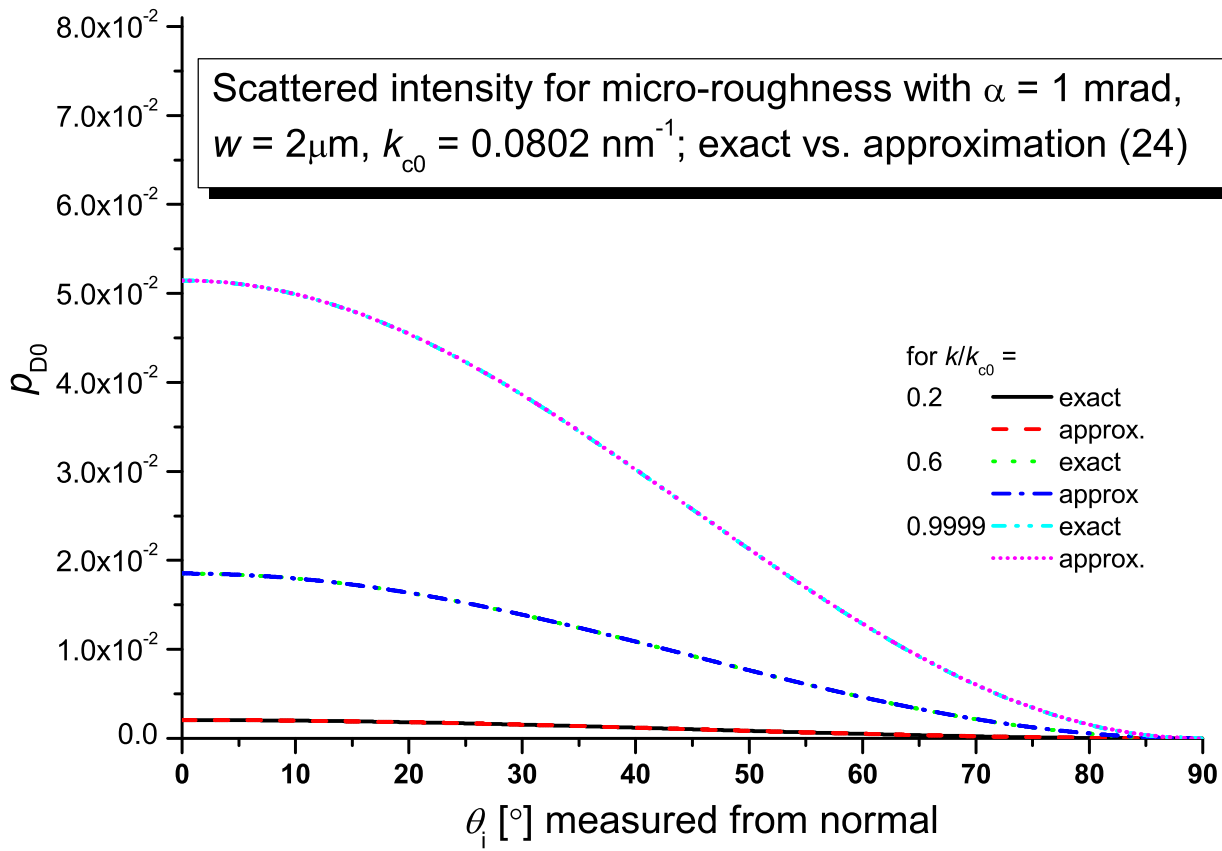}
\caption{(color online) The total diffusely scattered intensity $p_{D0}$ (from (18)
for the $K_{0}$ model) is compared to approximation (24) which is
valid for fairly smooth roughness ($\alpha<< 1$). For the parameters
used the agreement is excellent.}
\end{figure}

On a general basis, for $\alpha^2<<1$ the approximations (24-25)
are valid over a wide range of $\theta_{i}$, but not near grazing
incidence $(\theta_{i}\rightarrow \pi/2)$ since in this range
$(1-\textit{s}_{i})\textit{k}\textit{w}$ is $\textit{not}>>1$.
Therefore the $\nu$-range for integration over the full $\psi$-range
from $-\pi$ to $+\pi$ (see Fig. 1) cannot be extended to $\infty$.

From (25) we deduce a, perhaps, unexpected result: over a
significant range of $\theta_{i}$, namely where the curly bracket is
positive (for $\zeta <<1$: $\theta_{i}<45^{\circ}$), the loss
coefficient for the rough surface, $\mu(\theta_{i})$, is somewhat
smaller than that for the flat wall, $\mu_{0}(\theta_{i})$, with a
relative difference of order $\alpha^2$. This contradicts a general
statement in [7], p. 185, while a decrease was predicted for a
certain model in [11]. The decrease found here for $\alpha <<1$ is
independent of the model and is plausible for a geometric-optics
picture, where geometric reflections for the local surface
orientation are incoherently superposed: For near-normal incidence
the mean angle for incidence on inclined patches of a rough surface
is always larger than for the flat surface, for which it is zero for
$\theta_{i}=0$, and therefore the factor
$\left(\textit{k}_{c0}^{2}-\textit{k}^{2}\cos^{2}(\theta_{i})\right)^{-1/2}$
in $\mu_0(\theta_{i})$ is reduced, on average. No reduction of
$\mu_{0}(\theta_{i})$ is seen for 'jagged' roughness with $\alpha$
of order 1.
\begin{figure}
\includegraphics[clip=true,width=7in,angle=0]{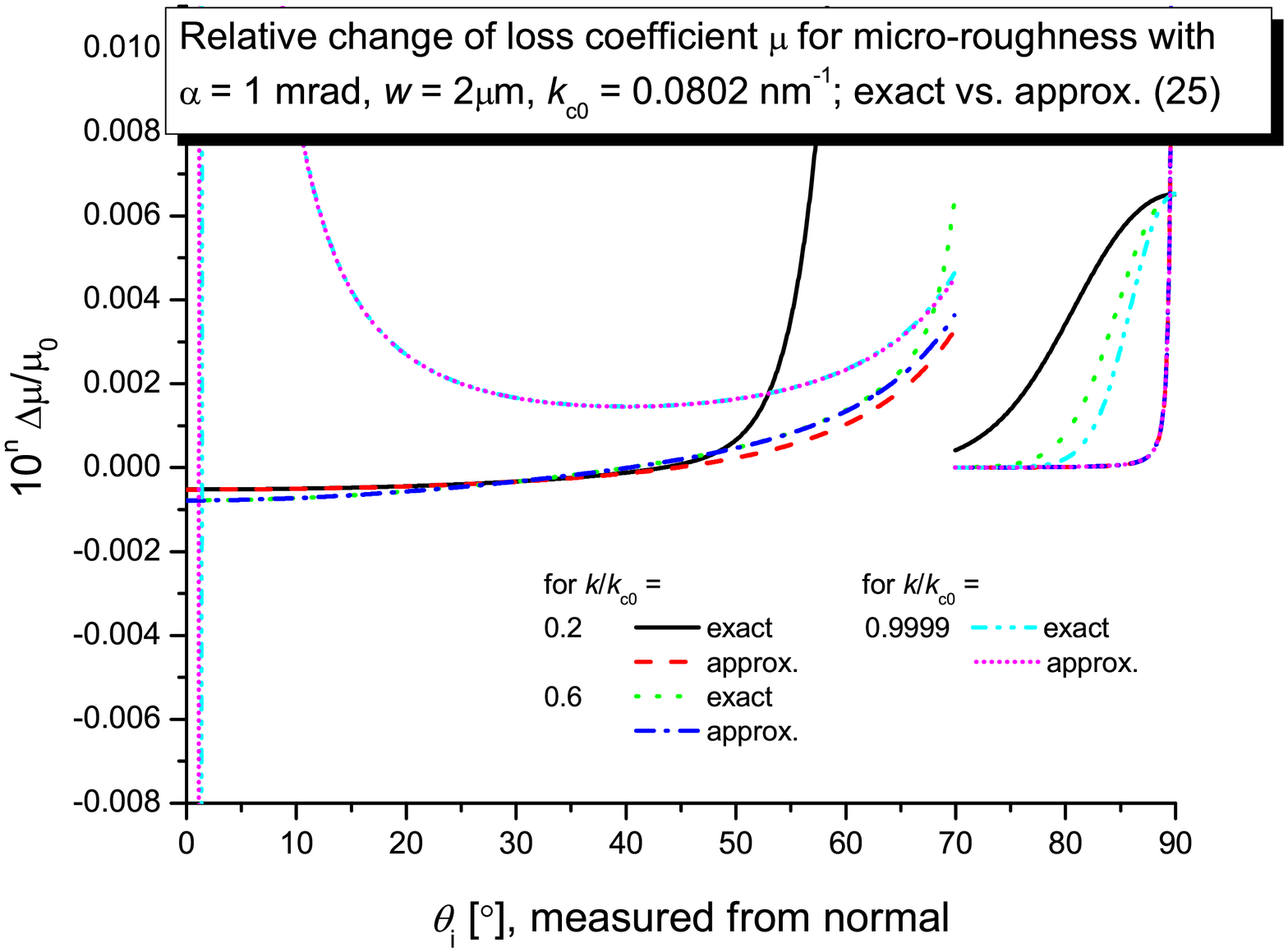}
\caption{(color online) Angle dependence of roughness correction to the wall loss
coefficient, $\Delta\mu(\theta_{i})/\mu_{0}(\theta_{i})$. The
numerical result from (20) for the $K_{0}$ model is compared to
approximation (25). The agreement is very good except for
$\theta_{i}$ near $90^{\circ}$ (glancing incidence). 
In this plot the scale changes from
 $n=3$ (expanded scale) for $\theta_{i}<70^{\circ}$ to
 $n=1$ (direct scale) for $\theta_{i}>70^{\circ}$.}
\end{figure}
For the same values of $\alpha$, etc. as for Fig. 2, Fig. 3 shows a
comparison of numerical results for
$\Delta\mu(\theta_{i})/\mu_{0}(\theta_{i})$ with approximation (25)
for the $K_{0}$ model. Again, the approximation clearly fails for
grazing incidence (both (24) and (25) diverge as
$\theta_{i}\rightarrow \pi/2)$ but is an excellent representation at
steeper incidence. The Gauss and $K_0$ model give very similar
results.

For UCN confined in a trap the angular distribution is, in most cases,
assumed to be close to isotropic (although deviations are critical,
see below). To obtain directional averages of $\textit{p}_{D0}$ and
$\Delta\mu$ the region of grazing incidence that is not described by
(24-25) has to be included, and it makes a significant contribution. The angular dependence of
$\Delta\mu(\theta_{i})/\mu_{0}(\theta_{i})$ for the Gauss-model (not
shown in Figs. 2 and 3) with the same parameters is close to that for the
$K_{0}$-model and the isotropic mean values
\begin{equation}
\left<Z\right>=2 \int^1_{0} c_{i}dc_{i} Z(\theta_{i})
\end{equation}
agree within 15 percent. Fig. 4 shows the 
$\zeta$-dependence of $\left<\Delta\mu\right>/\left<\mu_{0}\right>$
for both models. The reference average for the flat wall is given as [7-9]

\begin{equation}
\left< \mu_0 \right>
=2\eta[\arcsin\zeta-\zeta(1-\zeta^2)^{1/2}]/\zeta^2\rightarrow \left \{
\begin{array}{rcl} 4\eta \zeta/3\hspace{.1in}&for& \hspace{.1in}\zeta \ll 1\\ 
\pi\eta \hspace{.2in}& for &  \hspace{.2in} \zeta=1
\end{array}
\right.
\end{equation}

\begin{figure}
\includegraphics[clip=true,width=7in,angle=0]{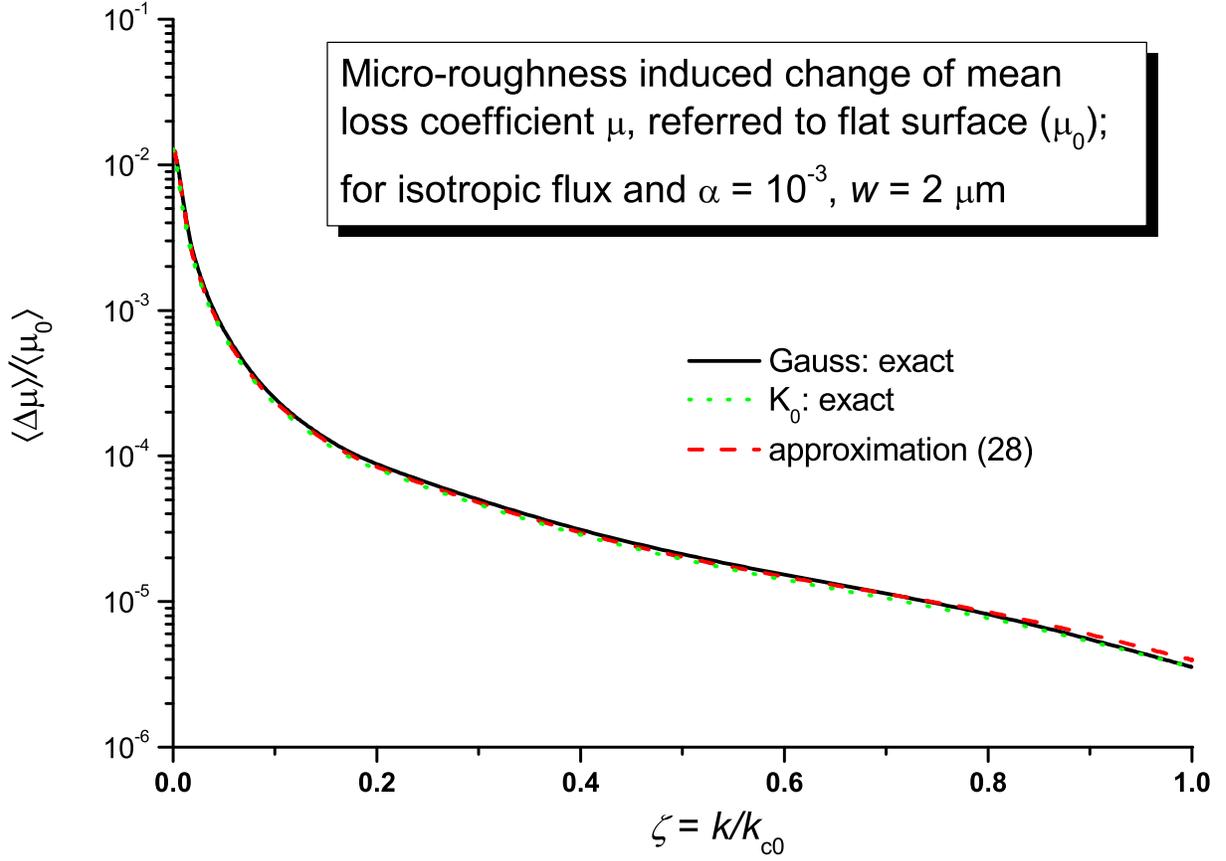}
\caption{(color online) Mean value of roughness correction,
$\left<\Delta\mu(\theta_{i})\right>$, for isotropic UCN flux,
referred to the flat-wall average $\left<\mu_{0}(\theta_{i})\right>$
given in (27). For the parameters given, the numerical result for
both models are compared to the semi-analytical approximation
(28). Both for the Gauss and the $K_{0}$ model, the agreement is
reasonably good, with maximal deviations of $\sim20\%$ over a wide
range of parameters.}
\end{figure}

The main features of $\left<\Delta \mu\right>/\left<\mu_{0}\right>$
are a sharp peak of height $(\textit{b}\textit{k}_{c0})^2$ at
$\zeta=0$ with half-width $\Delta\zeta\sim
\sqrt{2}/(\textit{w}\textit{k}_{c0})$ and a drop-off approximately
$\sim\zeta^{-3/2}$. A reasonable fit, with maximum deviations of
$\sim 20\%$ over a wide range of $\textit{k}_{c0}$ and of (small)
$\alpha$ and $\textit{b}$ and (large) $\textit{w}$, both to the
Gauss- and $K_{0}$ micro-roughness models, is
\begin{equation}
\frac {\left<\Delta\mu\right>}{\left<\mu_{0}\right>}\cong
k_{c0}^{2}b^{2}\left[1+3(\zeta k_{c0}
w/2)^{2}\right]^{-3/4}\left(1-\frac{1}{2}\zeta^{4}\right)
\end{equation}
which is shown as the dashed curve in Fig. 4.
   The isotropic-flux average of diffuse fraction $\textit{p}_{D0}$ is
\begin{equation}
\left<p_{D0}\right>=2k^2 b^2
\end{equation}
where we used the one-term approximation of (24).

   In applications to actual UCN
storage, approximations (28,29) can be used to determine further
averages analytically. For a monochromatic UCN spectrum under
gravity the averaging is over wall-interaction height \textit{h}
from the trap bottom to the 'roof' or to the maximum height
reachable for a given energy ($\textit{h}_{0}$, in units of maximum
jump height). Further averaging over the spectrum (over
$\textit{h}_{0}$) is usually also required. For instance, the sharp
peak of $\left<\Delta
\mu(\theta_{i})\right>/\left<\mu_{0}(\theta_{i})\right>$ at
$\zeta=0$ implies a large loss rate for UCN in contact with the wall
near their maximum jump height $\textit{h}_{0}$ where $\zeta
\rightarrow 0$. Taking into account that the flux
[$\sim(\textit{h}_{0}-\textit{h})\textit{d}\textit{S}$] incident on
surface element \textit{dS} vanishes as $\textit{h}\rightarrow
\textit{h}_{0}$, a detailed analysis is required. It shows that the
loss enhancement strongly depends on the trap geometry and UCN
energy and is significant for very low energy UCN 'hopping' in
small jumps on an essentially flat horizontal surface, for instance
the rim of a cylinder with horizontal axis, as in [3].

\section{\label{sec:level6}Detailed balance requirement and simplifications}

A trapped UCN gas will acquire or maintain an equilibrium isotropic
distribution only if its interaction with a rough wall (or magnetic
field irregularities) satisfies the detailed balance requirement
discussed in [7], p. 96. The flux undergoing scattering from solid
angle $\Omega_{i}=(\theta_{i},\varphi_{i})$ to
$\Omega=(\theta,\varphi)$ must equal the flux for the reverse
process $\Omega\rightarrow\Omega_{i}$, in accordance with the
fundamental principles of time-reversal invariance and
micro-reversibility. Since the flux incident at $\Omega_{i}$ (or
$\Omega$) is proportional to $\cos\theta_{i}$ (respectively
$\cos\theta$) any diffuse scattering distribution
$\textit{I}_{sc}(\Omega_{i}\rightarrow\Omega$) must satisfy the
symmetry requirement
\begin{equation}
I_{sc}(\Omega_{i}\rightarrow\Omega)=\tilde{I}(\Omega_{i},\Omega)\cos\theta
\end{equation}
where $\tilde{I}(\Omega_{i},\Omega)$ is a function symmetric in
$(\theta_{i},\theta)$ and $(\varphi_{i},\varphi)$ and $\cos\theta$
is the Lambert factor. The scattering distribution
$\textit{I}_{(11)}$ of Eq. (14) satisfies this requirement.

Various simplifications of the micro-roughness scattering distribution (14) (for
$\eta=0$) have been used in [13-15]. The limit of maximal diffusivity is reached
for a "dense roughness" model where 
$\textit{kw}\rightarrow 0$, hence $\alpha \sim b/w \rightarrow \infty$. This leads to [14,15]
\begin{equation}
I_{sc}\cong B \left [ \cos \theta_i\cos \theta \right ]\cos \theta
\end{equation}
with $B=2 k^4 b^2 w^2/\pi$ and a diffuse fraction
 $p_{D0}=(2\pi/3) B \cos \theta_i$
 that depends on incident angle $\theta_i$ and UCN energy
($\sim k^2$). It has been 
pointed out in [15] that for $kb$ at most of order 1, as required for
perturbation theory to be 
valid [4], the limit $kw\rightarrow 0$, hence $B\rightarrow 0$, is strictly
justifiable only for 
very small diffuse fraction ${p}_{D0}$.

As a further simplification, averaging over incident energy may be performed but
could be a coarse
 approximation for broad spectra, as for confined UCN. To arrive at the simplest
 form 
of (30), with $\tilde{I}=\textrm{const.}$, requires further averaging of
 $[\cos \theta_i \cos \theta]$ in (31). Assuming isotropic distributions in
$\theta_{i}$ and $\theta$ we obtain
$\left< \cos\theta_i \cos \theta \right> =4/9$.
 Thus, (31) becomes
\begin{equation}
\left< \!I_{sc} \right> =\left<\!\tilde{I}\!\right> \cos\theta
\end{equation}
with $\left<\!\tilde{I}\!\right>=8 \left< k^4 \right> b^2 w^2/(9\pi)$ and a 
total diffuse fraction $\left< p_{D0} \right> =\pi \left<\!\tilde{I}\right>$. 
Like (31), Eq. (32) is justified only for small $\left< p_{D0}\right>$.

In the "dense roughness" limit the loss coefficient $\mu (\theta_i)$ 
of (20) can be averaged analytically [7] with the result
 $\left< \mu \right>\, \cong\ \left< \mu_0 \right> (1+k_{c0}^2 b^2)$, 
where $\left< \mu_0 \right>$ is given in (27). However, except for very 
small $\left< p_{D0} \right>$ the correction term 
$k_{c0}^2b^2 \cong k_{c0}^4 \left< p_{D0} \right> /(\left<k^4 \right> w^2)$ 
would become large as $k_{c0}w\rightarrow 0$, conflicting with the
perturbation theory requirement.

It should also be noted that in averaging over $\theta_i$ we lose the 
proportionality of $I_{sc}$ to $\cos\theta_i$ which implies a small
scattering 
probability for glancing incidence ($\theta_i \rightarrow \pi/2$). 
For UCN stored in traps with high geometrical symmetry this dependence can 
give rise to almost stationary orbits for UCN "sliding" along a concave surface
 (see Sec. 8). For other geometries, only a few reflections, or where these details
are not important, as in the computer code tests of [13], (32) is a useful, 
simple approximation satisfying detailed balance.

As an example of a roughness model that does not satisfy detailed balance we mention 
the scattering distribution (E6) derived in Appendix E for the macroscopic, ray
 optics limit. In this case, ${I}_{sc}$ is symmetrical in
 $(\theta_i,\theta)$ 
and $(\varphi_i,\varphi)$, lacking the extra factor $\cos\theta$. This 
violation of detailed balance can be tolerated if
only a few reflections have to be considered, as in a neutron guide
[4], but attempting to use this model for thousands of consecutive
reflections in a Monte Carlo simulation, as required for long UCN storage
 (see below), results in the loss of isotropic equilibrium. We observed that UCN
accumulated in grazing-angle orbits, depleting other regions of phase
space.

\section{\label{sec:level7}Monte Carlo simulation techniques}

In Monte Carlo simulations of UCN propagation and storage, as in
[7,14,15], two complementary ways of implementing a given scattering
probability distribution $\textit{P}(\theta,\phi) $, e.g. Eq. (14),
have been discussed and used.

One ([7], p. 98) is based on mapping the joint probability
distribution $\textit{P}(\theta,\varphi)$  onto the ranges 0 to 1 of
uniformly distributed random numbers (r.n.) \textit{x},\textit{ y},
\textit{z}, etc. For instance, for the 'dense roughness'
limit of Eq. (14) used in [14] in the form
$\textit{P}(\theta,\varphi)\sim \cos\theta_{i}\cos^2\theta$, we can
determine $\theta$ and $\varphi$ from \textit{x}, \textit{y},
setting $\cos^3\theta=x $ and $\varphi/(2\pi)=y-\frac{1}{2}$. In
general, the mapping procedure involves indefinite integrals of
$\textit{P}(\theta,\phi) $ and becomes very tedious if numerical
integrations are required as for (14). In the simple case
$\textit{P}\sim \cos^2\theta$ the indefinite integral  $\int
\cos^2\theta d(\cos\theta)$ is $\sim\cos^3\theta$, hence the
prescription $\cos^3\theta=x$. Rendering this method efficient
requires 3D tabulation where the tables need only be interpolated at
each UCN reflection during a simulation run. As a further
simplification, once $\theta$ has been determined, $\varphi$ is
found faster from the conditional probability of $\varphi$, given
$\theta$:
$\textit{p}(\varphi)=\textit{P}(\theta,\varphi)/\textit{P}_{\theta}(\theta)$,
where $\textit{P}_{\theta}(\theta)$ is the integral probability for
$\theta$. For the Gaussian model the numerical work involves only the
incomplete Bessel function, which is readily available in computer code.

For our simulation (below) we chose the second, often faster method used
in [15], to determine $\theta$ and $\phi$ in a way consistent with
the probability distribution. First read from a table the total
diffuse-scattering probability $\textit{p}_{D0}$  for given
incidence parameters $\textit{k}$ and $\theta_{i}$. A r.n.
$\textit{x}_{0}$ then determines whether the reflection is specular
(if $\textit{x}_{0}>\textit{p}_{D0}$) or diffuse
($\textit{x}_{0}\leq\textit{p}_{D0}$). If diffuse, a trial set
$(\theta,\varphi)$ is obtained as $\cos\theta=\textit{x}$ and
$\varphi =2\pi (\textit{y}-\frac{1}{2})$. Next, find the probability
$\textit{P}(\theta,\varphi)$ for the trial set and compare it to a
fourth r.n., \textit{z}. The angles $(\theta,\varphi)$ are accepted
if $\textit{z}\leq \textit{P}(\theta,\varphi)/\textit{P}_{max}$,
where $\textit{P}_{max}$ is the maximum of
$\textit{P}(\theta,\varphi)$ for incidence at \textit{k} and
$\theta_{i}$. If
$\textit{z}>\textit{P}(\theta,\varphi)/\textit{P}_{max}$, the
process is repeated with new r.n.'s
$\textit{x}$,$\textit{y}$,$\textit{z}$. This scheme can be greatly
abbreviated if $2\pi\textit{P}_{max}<1$. In this case rolling three dice
$(\textit{x},\textit{y},\textit{z})$ only once is sufficient.
$\textit{x}$ and $\textit{y}$ determine $\theta$ and $\varphi$, as
before. If $\textit{z}>\textit{p}_{D0}$, the reflection is specular,
otherwise diffuse at angles $(\theta,\varphi)$. It can be verified
that this faster scheme also respects the probability distributions
correctly (as long as $2\pi\textit{P}_{max}<1$).

In all these schemes, wall losses and beta-decay are irrelevant for
the choice of reflection angles. However, we keep track of the
accumulation of net loss at each reflection and use the overall loss
factor as a weighting factor for UCN that have survived losses and
can be counted in a detector. The loss factor contained the second-order correction of Eq. (20) and was implemented in tabulated form.

\section{\label{sec:level8}Monte Carlo simulation for neutron lifetime experiment [3]}

The neutron lifetime experiment [3] reports a value
$\tau_{n}=878.5\pm 0.7_{stat}\pm 0.3_{sys}$ s which is 7.2 s (or
$0.8\%$, or $ >6$ standard deviations) away from the current
particle data world average $885.7\pm 0.8$ s [2]. The quoted
precision is better than for any previous single lifetime
measurement. Since data interpretation for this experiment relied
heavily on computer simulations we have performed independent
simulations using the roughness model outlined above. In this system
[16,3], shown in Fig. 5, a cylindrical or a 'quasi-spherical'
vessel with an opening rotates about a horizontal axis during the
various steps of a cycle: Filling with UCN with the hole pointing
straight down; rotation to a 'monitoring position' for spectral
cleaning; storage for a 'short' time (300 s) or a 'long time'
(2000 s) with the hole in the upright position; then emptying in 5
steps at intermediary angles. This scheme provides for a spectral
analysis, with a batch of higher-energy UCN counted first and the
slowest UCN pouring out at the last stage while the opening moves to
the vertical down position. The detailed measuring scheme [3] is
complex, involving not only energy dependent data for the 5 energy
bins but also traps with different mean free path for the UCN. This
combination allows, both 'energy extrapolation' and 'size
extrapolation' to the neutron lifetime.

\begin{figure}
\includegraphics[clip=true,width=3.4in,angle=0]{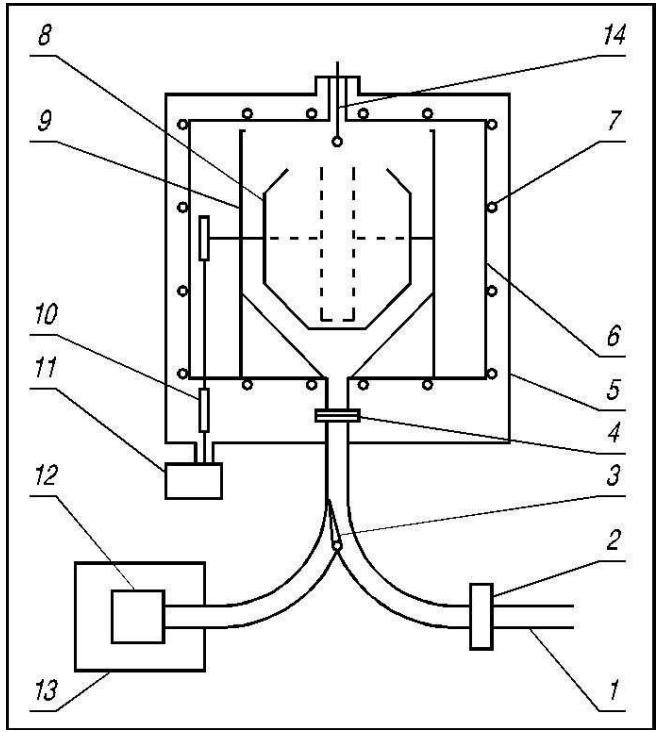}
\caption{Schematic of the apparatus used in [3]. UCN from a guide
tube 1 pass through a large entrance/exit channel to a cylindrical
or 'quasi-spherical' storage chamber 8 that can be rotated about a
horizontal axis.  After a short or long holding time, UCN in five
spectral ranges are, successively, discharged through the same
channel to the detector 12. The system 14 is used to coat the walls
with a special low-temperature oil.}
\end{figure}

Data analysis is also complex, relying heavily on computer
simulation to determine the parameter $\left<\gamma\right> =
\left<\mu\textit{f}\right>/\eta$ against which measured inverse
storage lifetimes are plotted for the extrapolation. The average
$\left<\mu\textit{f}\right>$, with wall collision frequency
\textit{f}, is the wall loss rate (mean loss per second).

For a narrow cylindrical trap the authors of [3] compared simulations with the
measured time spectra for the first counting interval following short storage
(300s) (see Fig. 14, lower part, of [3]). They concluded that a certain minimum
roughness (diffuse fraction $\geq 1 \%$) of the trap wall was required to
obtain agreement. This is a crucial step with respect to the neutron lifetime,
and therefore we have performed independent simulations of short and long
storage cycles for a narrow cylindrical vessel with radius 38cm, width 14cm, and
aperture ${62}^{\circ}$ for the opening, to investigate this issue using the
near-mirror reflection 
model with long-range roughness ($\alpha<<1$) described above.

In our simulation UCN are generated at the beginning of the
monitoring phase (at trap angle $30^{\circ}$) over a horizontal area
inside the trap volume and near its bottom. The distribution is
isotropic with polar angle $\theta$, referred to the up and down
directions, derived from a r.n. $\textit{x}$ as
$\cos\theta=\textit{x}^{1/2}$. The energy distribution is the same
as used in [3], namely the Maxwell distribution modified by a
heuristic attenuation factor $\exp(-\textit{h}_{0}/\textit{h}_{sp})$
with $\textit{h}_{sp}=0.6$ m (not specified in [3]). We used a
maximum (minimum) energy of 0.95 m (0.05 m).

The time sequence of monitoring, storage and emptying in five steps
was the same as described in [3]. The wall reflections were assumed
elastic except for the (very small) Doppler shift for wall
reflections during times when the trap rotates at 
$5.1^{\circ}/$s (respectively $17^{\circ}/$s 
on the way from $0^{\circ}$ to $40^{\circ}$ for a rotation time of 2.3s [3]) 
and the reflection is diffuse. For perfectly cylindrical geometry the 
overall shift is very small. Reflection points were calculated
analytically from the previous reflection position, energy and
take-off angle, assuming perfectly cylindrical geometry.  At each
reflection, the wall loss was incremented according to the loss
coefficient $\mu$ for this reflection. A run was terminated at the
time when the UCN passed through the opening without immediately
falling back into the trap. For the purpose of this work the exit
channel could be neglected and those UCN that had escaped loss, as
determined by the integrated loss including beta-decay, were counted
into time bins of 10s or 1s resolution. For consistency with [3] we used
the quoted values $\eta=2\times10^{-6}$ and $\tau_{n}=878.5$ s in
the simulation. The bins were combined to the five counting
intervals, and short runs with storage time 300 s were compared to
long runs (2000 s) as in the actual experiment to obtain the storage
lifetime as a function of mean values $\left<\gamma\right>$ that
were also calculated for each counting interval.

\begin{figure}
\includegraphics[clip=true,width=6in,angle=0]{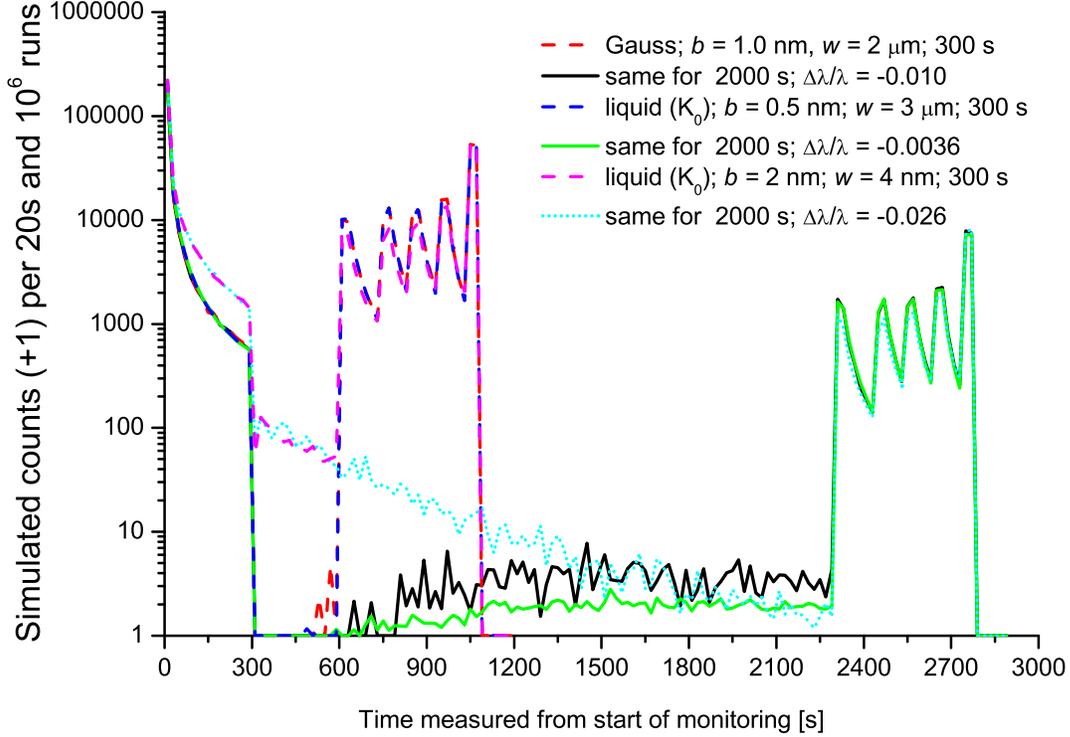}
\caption{(color online) Simulated count rates, in 20 s bins, versus cycle time
\textit{t}, starting from 'monitoring' for 300 s where the trap is
at hold position $30^{\circ}$ titled away from vertical. The data
are normalized  to $10^{6}$ runs (one run for each UCN started at
\textit{t} = 0). Data are shown for both short and long storage and
for both the Gauss and $K_{0}$ models for a fairly smooth surface
with similar parameters, as shown in the inset.  They correspond to
$\alpha=7.1\times10^{-4}$ for the Gauss-model, and
$\alpha=2.4\times10^{-4}$ for $K_{0}$. An example of steeper roughness
($\alpha=0.7$) is shown for comparison. We plot the simulated counts
N \textit{plus one} to allow a distinction between N = 0 and N = 1
on the log scale. In the simulations of Fig. 6 (only) we used a constant 
angular speed of $4.3^{\circ}/$s for all trap rotations.}
\end{figure}

Figure 6 shows simulated count rates, in 20 s bins, versus cycle
time $\textit{t}$, starting from 'monitoring' for 300 s where the
trap is at position $30^{\circ}$ tilted away from vertical. The data
are normalized to $10^{6}$ runs (one run for each UCN started at
\textit{t} = 0). Data are shown for both short and long storage and
for both the Gauss and $K_{0}$ models, for a fairly smooth surface
with similar parameters, as shown in the inset. They correspond to
$\alpha=7.1\times 10^{-4}$ for the Gauss-model, and $\alpha
=2.4\times10^{-4}$ for $K_{0}$. An example of steeper roughness with
$\alpha = 0.7$ (for $\textit{b}=2$nm and $\textit{w}=4$nm) is shown for comparison.

We will focus on two features of Fig. 6, and their implications:
 first the time spectra for the 
first counting interval following short storage; then  the count-rates 
during long storage.

\vspace{10pt}

\textbf{First counting peak}

To allow a comparison with the time spectra for the first peak measured in [3],
we show in Fig. 7, at a higher resolution of 1 to 2s, an expanded view of the
interval from 600 to 750s (on the time scale of Fig. 14 of [3] this corresponds
to the interval 740 to 890s). The data of [3] are included in Fig. 7 as curves 1
and 2. Prior to discussion we point 
out differences between the two simulation approaches.

\begin{figure}
  \begin{center}
    \begin{tabular}{c}
      \resizebox{5in}{!}{\includegraphics{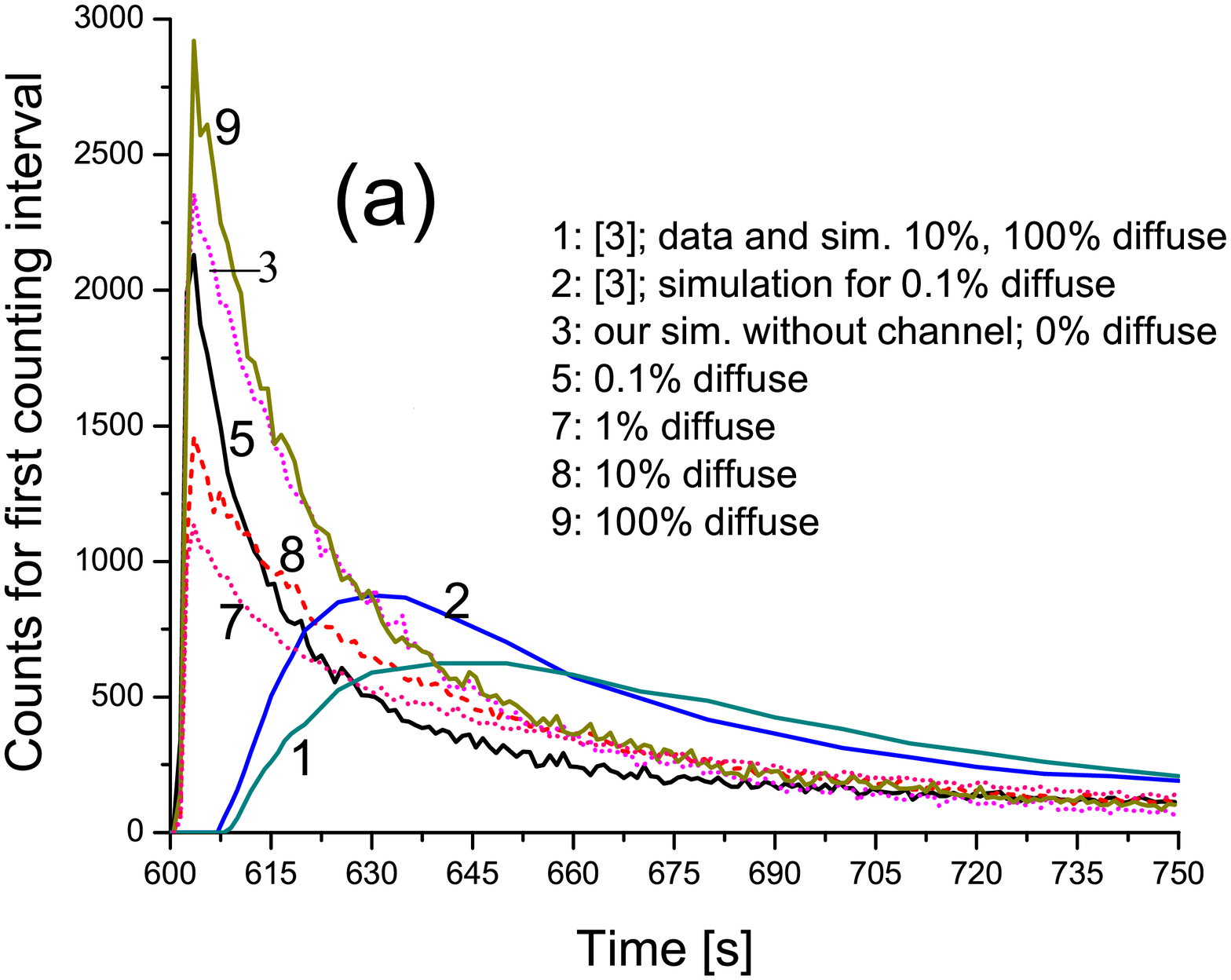}} \\
      \resizebox{5in}{!}{\includegraphics{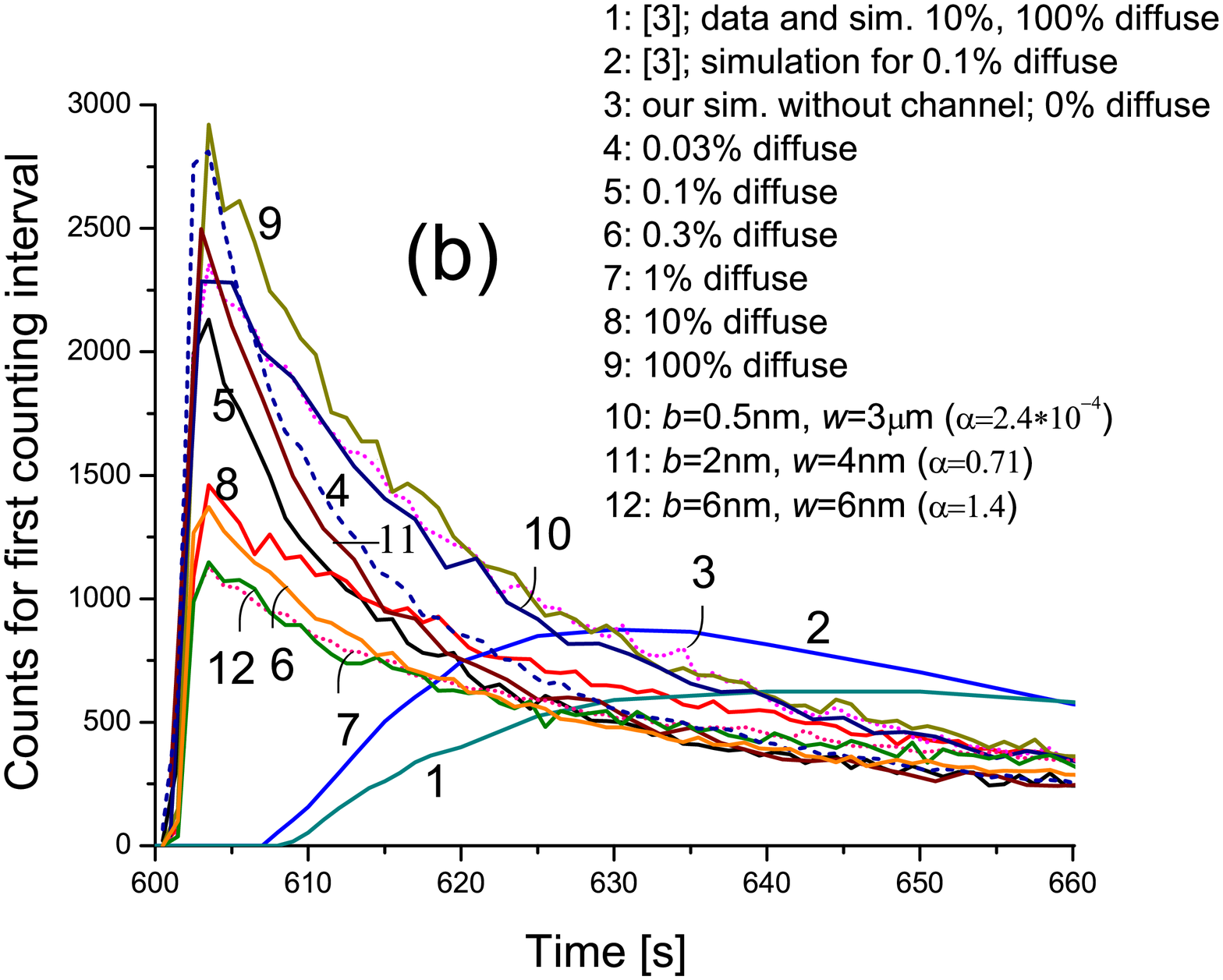}} \\
    \end{tabular}
    \caption{(color online) Comparison of measured [3] and simulated time spectra for the first
  counting period following short storage. For clarity, part (b) shows an
  expanded view of the initial 60s of the full 150s interval shown in
  part(a). Some of the simulation curves of part (b) 
are also included in (a). For a discussion see the text.}

  \end{center}
\end{figure}

The authors of Ref. [3] used a roughness model characterized by a single number,
the diffuse fraction for the trap walls, but did not state which roughness model
was used. The simulation included the exit channel (designated secondary volume
and UCN guide in [3]) but the authors did not describe the geometrical model
used for this complex structure. It included (see Fig. 5) the trap exterior and
interior (which some UCN temporarily re-enter on their way to the detector), the
wide conical channel section, the valve selector with its gaps, the curved guide
and the detector with its window that accumulated an oil film as a result of
repeated surface coating using the evaporator 14 shown in Fig. 5. The surface
properties, including roughness, of the oil-coated channel walls with their
temperature gradient, varying from the trap 
temperature to near-room temperature for the detector, are not discussed.

Our simulation did not include the exit channel. We used the roughness models
and parameters listed for curves 3-12 of Fig. 7. In Fig. 7b the initial part is
expanded for clarity. For these counting spectra following short storage the
details of wall loss are insignificant and the results using (20) are
indistinguishable from those for the elementary, flat-wall expression that was
used in [3]. (During the long holding period of 2000s the details of 
loss coefficient may become important, depending on the roughness parameters.)

Curve 1 represents the measured data (Fig. \!14 of [3]), using a scale
approximately adjusted to counts per 2s for $10^{6}$ initial UCN. Curve 1
practically coincides with the Ref. [3] simulations for 10\% and 100\%
diffusivity. Curve 2 is the simulation, in [3], for 0.1\% diffusivity. Curve 3
is our simulation (for the trap only) for an ideally smooth surface. For curves
4-9 we used the simplest scattering distribution consistent with detailed
balance, Eq. (32), with diffuse fractions 0.03, 0.1, 0.3, 1, 10 and 100\%. For
consistency with [3], we included large values of $\left<p_{D0}\right>$,
disregarding the restriction outlined following Eq. (31). Curves 10-12 are based
on the micro-roughness result (14) (for $\eta=0$) and the
$K_0$-model. The parameters for curves 10-12 represent increasingly
steeper roughness ($\alpha=2.4\times 10^{-4}$, 0.71, and 1.4) and increasing
$\left<\textit{p}_{D0}\right>$ 
(0.16, 0.55, and 11.3\%, where the last value is at the limit of perturbation
 theory validity).

Fig. 7 shows the large spectral distortion and time delay due to the exit
channel. The peak widths at half maximum for curves 1 and 2 are 3 to 6 times
larger than for the simulations excluding the exit channel. Thus the channel
response function to the relatively short pulse 
of UCN pouring out of the trap essentially determined the measured time spectra.

For consistency with [3] we focus first on the simulations based on (32). The
minimum half-width $\Delta\textit{t}$=10s is observed for
$\left<\textit{p}_{D0}\right>$=0.1\%. Both for
$\left<\textit{p}_{D0}\right>\geq$1\% and $\left<\textit{p}_{D0}\right><<0.1\%$
(including zero), the curves are $\sim$2 times wider, while the curves for
0.03\% and 0.3\% are intermediary. The difference between 0.1\% and $\geq$1\%
diffusivity makes plausible the (much smaller) difference of broadened peaks
between the Ref. [3] simulations (including the channel) for
$\left<\textit{p}_{D0}\right>$=0.1\% (curve 2) and 10 or 100\% (curve
1). Comparing to the measurement the authors rejected the possibility of
diffusivity $<<1\%$ for the trap. However, curve 3, for
$\left<\textit{p}_{D0}\right>$=0, 
has about the same larger width, namely
 $\Delta\textit{t}$=23s, as for $\left<\textit{p}_{D0}\right>\geq$1\% (curves 7-9).

The same type of broad line is seen for curve 10 that is based on Eq. (18) for 
small $\alpha$. The intermediate (steep) roughness model used for 
curve 11 (12) results in intermediary (broad-type) linewidths.

The difficulty of distinguishing between roughness models for the trap on the
basis of measured curves broadened 3 to 6 times by the response, or resolution,
function is exacerbated by two aspects: First, by the uncertainty introduced by
any simplification of the complex channel geometry, as needed to make the
simulations feasible. Secondly, in this experiment (and perhaps any UCN
experiment conducted so far) neither the primary source intensity and spectrum
nor the transmission characteristics of connecting guide tubes, shutters and
other components, here including the channel that is passed by the UCN also
during trap loading, is known well in absolute terms. Thus, the area under the
counting peak can practically not be used as a criterion for adoption or
rejection of a roughness model for the trap. In the present case, the observed
variation of peak area in Fig. 7 is a 
reflection of spectral mixing between the five spectral intervals, as discussed below.

We conclude from this peak analysis that a "soft-roughness" model with
$\alpha << 1$, as is expected for the temperature-cycled [3] oil-glass coating
used, can, 
probably, not be excluded. This has the consequences outlined below.

\vspace{10pt}

$\textbf{'Spill-over' during long storage}$

Returning to Fig. 6, we draw special attention to the finite count rate that appears
during the storage intervals, $\textit{t}$ = 300 to 600 s
for short storage and $\textit{t}$ = 300 to 2300 s for long storage.
At the end of monitoring at trap angle $30^{\circ}$ most UCN with
energies exceeding the height barrier ($\sim56.5$ cm above trap
bottom) have spilled out. However, some higher energy UCN move,
preferentially, along the cylinder axis and little in other
directions. In this highly symmetric trap geometry they have
survived in nearly stationary orbits since, for $\alpha<<1$, the
reflections are nearly mirror-like with small reorientations of
order $\alpha$ if the reflection is diffuse. Thus the random walk
toward isotropy is very slow. As the trap rotates to angle
$0^{\circ}$ at $\textit{t}=300$ s the barrier rises by $\sim $ 14 cm
to 70.5 cm from the bottom, and the count-rate in Fig. 6 drops to
zero. However, since UCN with energies exceeding this new barrier
still remain, the same slow relaxation toward isotropy gradually
results in steeper orbits. At $\textit{t} \sim 600$ s some UCN have
reached the new barrier and we see a fairly constant 'background'
count rate throughout the remainder of the long storage interval up
to $t = 2300$ s. The range of parameters $\alpha$, $\textit{b}$ and
$\textit{w}$ showing this general tendency is fairly broad, and
there is no qualitative difference between Gauss and $K_{0}$ models.
We will show that this 'background' could be overlooked in actual
measurements. The same slow diffusion in phase space is also
expected to mix the spectral regions for the five counting
intervals, and the net result of slow relaxation could give rise to
a significant systematic error.

For comparison with experiment [3] we note that the counts in Fig.
6, for $10^{6}$  runs, correspond to $\sim 12$ double cycles (one
cycle each for short and long storage). The instrument background of
$\sim $0.02 s$^{-1}$ (Fig. 10 of [3]) is equivalent to $\sim$ 5
counts/20 s in the region 600 s $<\textit{t}<$ 2300 s of extra
holding time for long storage in Fig. 6. This interval has no counts
without the relaxation process since the instrument background was
not included in our simulations. Thus, the UCN 'spilling over the
threshold' during this time ($\sim$ 0.8/20 s, or 67 in total for the
$K_{0}$ model) are missing from the UCN count, $\textit{N}_{long}$,
following long storage, but are not missing from the short-storage
counts $\textit{N}_{short}$. In the actual experiment the relative
change of measured 'background' (0.8/5 $\sim 15\%$) is below the
significant background variations shown in Fig. 10 of [3] and would
hardly be noticed. However, calculating from
$\textit{N}_{short}/\textit{N}_{long}$ an inverse storage lifetime,
$\lambda=1/\tau_{st}=
[\ln(\textit{N}_{short}/\textit{N}_{long})]/\Delta\textit{t}$, with
$\Delta\textit{t}=\textit{t}_{long}-\textit{t}_{short}$ = 2300 s
-600 s = 1700 s, disregards the 'spill-over' and would result in a
significant uncertainty of $\lambda$, as shown in the following
estimation.

We assume that the spill-over rate $n_{spill}(\textit{t})$
originates exclusively from the highest-energy interval $\# 1$ (of
the 5 intervals) and that entry into (exit from) spectral interval
$\#1$ by angular diffusion from (into) $\#2$, or any other interval,
is negligible. Thus the number $\textit{N}_{1}(\textit{t})$ of UCN
in $\#1$ decreases during long storage as
\begin{equation}
\frac{1}{N_{1}(t)}\frac{dN_{1}}{dt}=-\frac{1}{\tau_{st}}-\frac{1}{N_{1}(t)}n_{spill}(t)
\end{equation}
In [3] $\tau_{st}$ is close to the lifetime $\tau_{n}$,  so we can
approximate $N_{1}(t)$ by $N_{1} (t_{short})\exp[-(t - t_{short})/
\tau_{n})]$ in the last term of (33). Integrating (33) from
$t_{short}$ to $t_{long}$ and equating $N_{1}(t_{short})$ with the
counts for interval $\#1$, $N_{1,short}$, and $N_{1}(t_{long})$ with
$N_{1,long}$ we obtain
\begin{equation}
\lambda=\frac{1}{\tau_{st}}\cong \frac{1}{\Delta
t}\ln\frac{N_{1,short}}{N_{1,long}}-\frac{C_{spill}}{N_{1,short}\Delta
t}
\end{equation}
The first term on the right is the $\lambda$-value disregarding
spill-over. The second term is the correction with
\begin{equation}
C_{spill}=\sum_{i} n_{spill}(t_{i})\exp[(t_{i}-t_{short})/\tau_{n}]
\end{equation}
where the sum is over the count rates $n_{spill}(t_{i})$ from
$t_{short}$ to $t_{long}$, with a weight factor
$\exp[(t_{i}-t_{short})/ \tau_{n}]$. Finally, the correction becomes
\begin{equation}
\frac{\Delta \lambda}{\lambda}=-\frac{\Delta
\tau_{st}}{\tau_{st}}\cong
-\frac{\tau_{n}C_{spill}}{N_{1,short}\Delta t}
\end{equation}

This value is given in the inset of Fig. 6 and amounts to
significant corrections to the inverse storage lifetime for interval
$\#1$. If inter-interval diffusion is included the values for other
spectral intervals, and presumably the energy extrapolation as well,
are affected to a significant extent.  On the other hand, the
measuring scheme of [3] is more complex, involving also the larger
traps including a 'quasi-spherical' vessel, which in the
simulations was replaced by a cylinder in a way not discussed in
[3]. We will point out below that our simulation for a fairly smooth
surface of a narrow cylindrical trap leads to questions also
regarding the $\textit{x}$-values of the extrapolation, namely the
mean values $\left<\gamma\right>$ for the spectral intervals 1-5.

For the example with steeper roughness in Fig. 6, relaxation is faster
and the large 'back-ground' in the region 300 s $<$ t $<$ 600 s
might be visible in the experimental data, in spite of blurring by
the exit channel.

\begin{figure}
\includegraphics[clip=true,width=7in,angle=0]{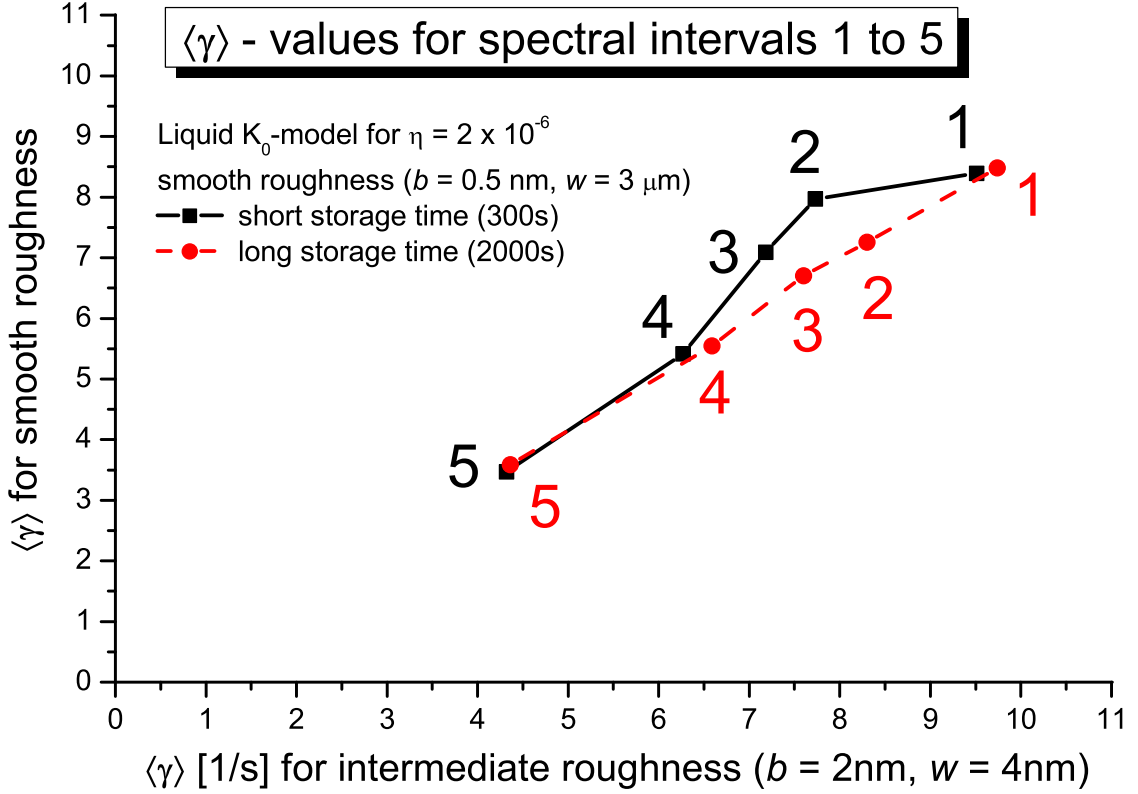}
\caption{(color online) Simulation results for the mean value $\left<\gamma\right>
= \left<\mu\right>/\eta$, for the five spectral ranges 1 to 5 of
[3], plotted for short storage (300 s, solid lines) and for long
storage (2000 s, dashed lines).  For the $K_{0}$ model, the abscissa
shows an example of 'intermediary roughness' with $b = 2$ nm, $w = 4$ nm,
thus $\alpha = 0.7$. The ordinate values are for $b = 0.5$ nm, $w =
3$ nm, thus $\alpha = 2.4\times10^{-4}$. The seemingly 'erratic'
behavior in either case may be explained by mixing between the five
spectral intervals and loss due to 'spill-over' during the storage
time as discussed in section 8. The statistical errors for the
$\left<\gamma\right>$ values are insignificantly small.}
\end{figure}

\vspace{10pt}

{\textbf {Mean}} $\boldsymbol{\gamma}${\textbf{-values}}

For the narrow cylindrical vessel, Fig. 8 combines
$\left<\gamma\right>$ values for the five spectral intervals: on the
\textit{x}-axis for an 'intermediate' roughness with $b = 2$ nm and $w =
4$ nm, and on the $\textit{ y}$-axis for a 'smooth' roughness with
$b = 0.5$ nm and $w = 3$ $\mu$m. The data for short and long storage
are plotted separately and the differences are seen to be
substantial. They are not explained by the small spectral cooling
during storage due to larger wall losses for higher UCN energies.
Spectral mixing and spill-over are more important. The large
differences are also seen in the mean collision frequencies (not
shown) for the five intervals.

\begin{figure}
\includegraphics[clip=true,width=7in,angle=0]{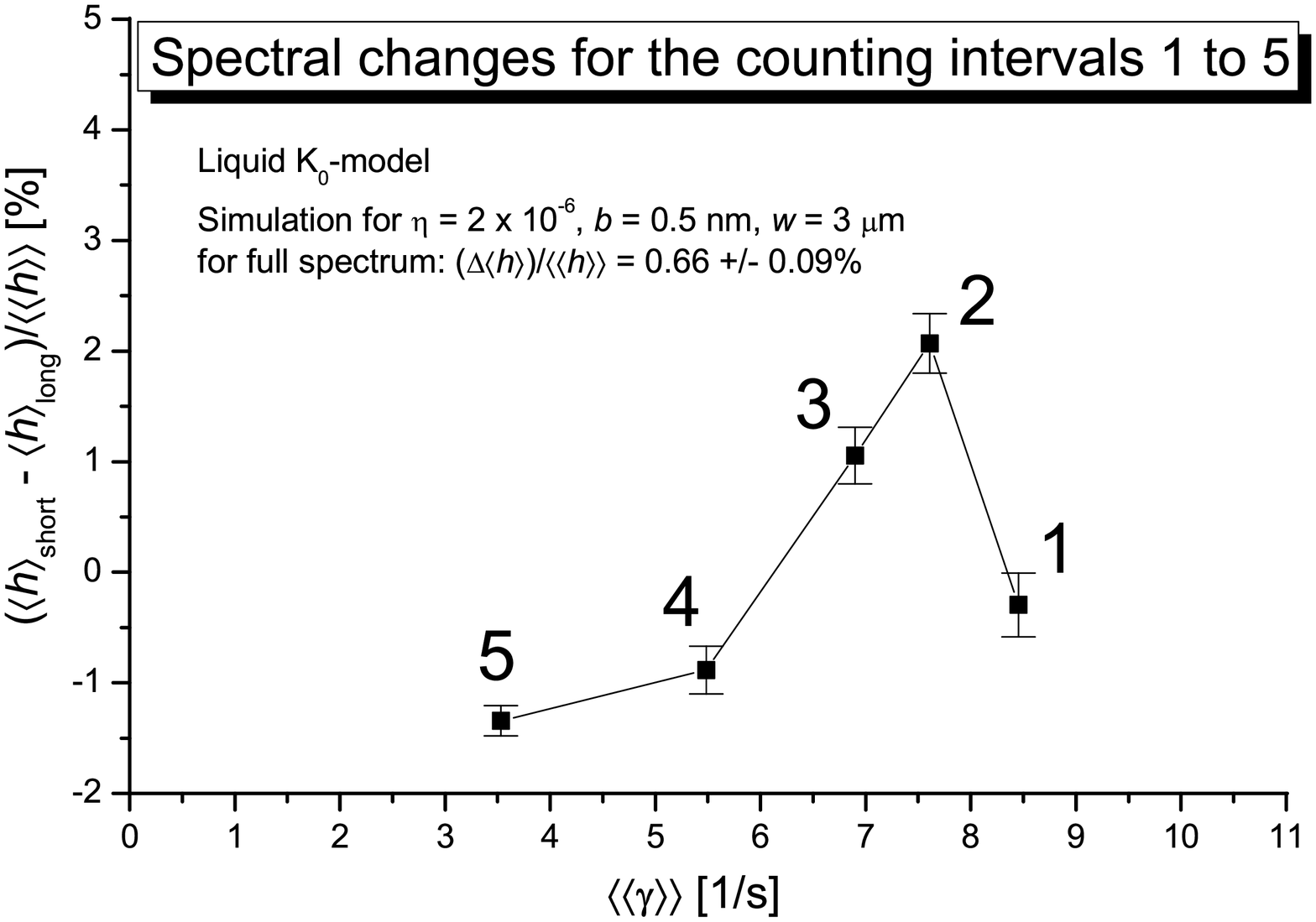}
\caption{The 'cooling' effect during long storage versus short
storage. Instead of a small systematic drop $\left<\Delta h\right>$
of mean spectral height $\left<h\right>$ as a function of time,
expected since the wall loss increases with UCN energy, the
simulation shows fairly large and 'erratic' values of $\left<\Delta
h\right>$ for the five spectral ranges. Although there is cooling
for the full spectrum, the lowest-energy intervals show negative
values of $\left<\Delta h\right>$ ('heating'). The simulation data
are for the $K_{0}$ model with fairly 'soft roughness' ($\alpha =
2.4\times10^{-4}$) and are explained by inter-spectral mixing and by
'spill-over' (see section 8). The
$\left<\left<\gamma\right>\right>$ and $\left<\left<h\right>\right>$
values are averages over the spectral ranges and over the two
storage times (300 s and 2000 s).}
\end{figure}

\vspace{10pt}

\textbf{Mean spectral energies}

Spectral mixing is also seen in the 'erratic behavior' of mean
spectral energies (height $\left<h\right>$ and difference,
$\Delta\left<h\right>$, between short and long storage). In Fig. 9,
$\Delta\left<h\right>/\left<\left<h\right>\right>$ is plotted versus
$\left<\left<\gamma\right>\right>$ (averaged over long- and short
storage) for the 'smooth' roughness parameters. A spectral change
between short and long storage implies a change of detection
efficiency that is not discussed in [3]. While the difference in
mean UCN velocity is reduced by the energy boost through the fall
height of $\sim1.5$ m between storage vessel and detector, the
transparency of the  oil coated detector window for an isotropic UCN
beam is a steep function of velocity in the region of interest, with
roughly a change of 4\% for 1\% of UCN energy change. As a result
the UCN spend more time in the entrance/exit channel and have a
greater chance of being lost, for instance through gaps.

\section{\label{sec:level9}Conclusions}

Using models of microscopic roughness characterized by height-height
correlation functions of the Gauss and '$K_{0}$' type (for solid
vs. liquid surfaces) we have carried perturbation calculations to
second order, recovering a previous result [7] for roughness-induced
wall loss for UCN that has, apparently, found no attention so far. It
may be surprising that under certain conditions roughness reduces
the wall loss probability. Furthermore, the second-order approach
provides absorption corrections to the diffuse scattering
distribution as well as the leading terms of a 'Debye-Waller factor
for roughness' that is consistent with previous derivations but does
not  require the additional assumption of a height probability
distribution function made, e.g., in [1] and [5]. It is shown that
for a fairly smooth wall with small mean-square slope the results
are model independent and are determined only by the mean values,
mainly for height ($b^{2}$) and slope ($\alpha^{2}$).

As the central application of roughness scattering we describe 
independent simulations relating to neutron
lifetime experiment [3] which reported a significant deviation from
the world average and very high precision. The possibility of a
fairly smooth surface for the temperature cycled oil used 
as the wall coating in this
experiment, in connection with the possibility of nearly stationary
orbits in any highly symmetric trap geometry, raises questions about
the reliability of the extrapolation method used to extract the
lifetime value from the storage data.

\begin{acknowledgments}
We acknowledge very useful discussions with B.~G.~Yerozolimsky,
G.~M\"{u}ller, R.~Golub, E.~Korobkina, A.~V.~Strelkov and V.~K.~Ignatovich.
We thank N.~Achiwa for measuring the temperature
dependence of the scattering potential for the LTF oil.
\end{acknowledgments}

\pagebreak

\appendix

\section{Correlation functions}
Application of the Laplace operator twice in Eq. (7) yields for the
Gauss model:
\begin{equation}
h(\delta)=\left<\kappa(\rho)\kappa(\rho+\delta)\right>=\kappa_{G}^{2}\left(1-\frac{\delta^{2}}{w^{2}}+\frac{\delta^{4}}{8w^{4}}\right)
\exp[-\delta^{2}/(2w^{2})]
\end{equation}
with mean square curvature
\begin{equation*}
\kappa_{G}^{2}=h(0)=\frac{\alpha^{2}}{w^{2}}=\frac{2b^{2}}{w^{4}}
\end{equation*}
Using properties of the modified Bessel functions $K_{n}$ (with
$n=0,1,2...$) we obtain for the $K_{0}$ model.
\begin{equation}
h(\delta)=\kappa_{K}^2\frac{Q(z)}{Q(t)}
\end{equation}
with
\begin{equation*}
z=[(\delta^{2}+\delta_{0}^{2})/(2w^{2})]^{1/2};\hspace{.2in}t=\delta_{0}/(w\sqrt{2})
\end{equation*}
\begin{equation}
Q(z)=8\frac{K_{2}(z)}{z^{2}}-\frac{\delta^{2}}{\delta^{2}+\delta_{0}^{2}}\left[8\frac{K_{3}(z)}{z}-\frac{\delta^{2}}{\delta^{2}+\delta_{0}^{2}}K_{4}(z)\right]
\end{equation}
and
\begin{equation}
\kappa_{K}^{2}=h(0)=\frac{1}{8}\frac{\alpha^{2}}{w^{2}}\left[\frac{K_{3}(t)}{K_{1}(t)}-1\right]
=2.103 \alpha^{2}/w^{2}
\end{equation}
The expressions for the mean squared curvature hold for the choice
$t=0.7709$, for which the two models have identical mean-square
slope $\alpha^2$. This value of $t$ is the solution of
$K_{2}(t)=5K_{0}(t)$.

\pagebreak

\section{Green's function for a flat surface}
Eq. (10) describes the wave, with wave number \textit{k}, generated
at point $\textbf{r} =(\boldsymbol\rho,z)$ by a point source at
$\textbf{r}'=(\boldsymbol\rho',z')$ near the surface $z=0$ of a
semi-infinite slab of material with scattering potential
$k_{c}^{2}/4\pi$. The solution may be represented by a Fourier
expansion in terms of in-plane waves
$\exp(i\textbf{k}_{\parallel}\cdot\boldsymbol\rho)$ as [17]
\begin{equation}
G(\boldsymbol\rho,z|\boldsymbol\rho',z')=\frac{1}{(2\pi)^{2}}\int
d^{2}\textbf{k}_{\parallel}e^{i\textbf{k}_{\parallel}\cdot(\boldsymbol\rho-\boldsymbol\rho')}g(z|z')
\end{equation}
The one-dimensional Green's function for $k_{z}$ satisfies the
equation
\begin{equation}
\frac{d^{2}g(z|z')}{dz^{2}}+K_{z}^{2}(z)g(z|z')=-4\pi\delta(z-z')
\end{equation}
where the wave number component perpendicular to the wall is
$k_{z}=(k^{2}-k_{\parallel}^{2})^{1/2}$, and $K_{z}=k_{z}$ in vacuum
and $K_{z}=k_{z}'=(k_{z}^{2}-k_{c}^{2})^{1/2}$ inside the medium.
For UCN with $k^{2}<k_{c0}^{2}$,
$k_{z}'=i\kappa=i(k_{c}^{2}-k_{z}^{2})^{1/2}$ is imaginary apart
from a very small real part due to the loss contribution $i\eta$ in
$k_{c}^{2}=k_{c0}^{2}(1-i\eta)$. The solution of (B2) for outgoing
plane waves can be written, for $z>z'$, $z>0$:
\begin{eqnarray}
  g(z|z')&=& \frac{4\pi i}{k_{z}+i\kappa} e^{ik_{z}z}e^{\kappa
z'}\nonumber\\
  &\cong&\frac{4\pi i}{k_{z}+i\kappa}e^{ik_{z}z}(1+\kappa z')
\end{eqnarray}
and for $z<z'$, $z<0$:
\begin{eqnarray}
  g(z|z')&=& \frac{2\pi}{\kappa}e^{\kappa z}(e^{-\kappa z'}-R e^{\kappa z'})\nonumber\\
  &\cong&\frac{4\pi i}{k_{z}+i\kappa}e^{\kappa z}(1+ik_{z}z')
\end{eqnarray}
with $R=(k_{z}-i\kappa) /(k_{z}+i\kappa)$.

 (B3) and (B4) were
derived assuming $z'<0$. The two-term approximations in (B3) and
(B4) are valid for 'micro-roughness', where $k_{c0}b<<1$, and they
hold for positive or negative values of $z'$ within the rough layer.
This is a direct consequence of the continuity of $\psi$ and
$d\psi/dz$ at the interface. We note that the first two expansion
terms are sufficient for a full description of UCN interaction with
a rough wall up to terms quadratic in the roughness amplitude
\textit{b}.

The same kind of approximation can be made for the plane-surface
wave $\psi_{0}(\textbf{r}')$ in the integral of Eq. (9). Thus, using
(B3) the $z'$-integral in (9) from $z'=0$ up to the roughness
amplitude $\xi(\boldsymbol\rho')$ across the rough layer is
proportional to $\xi+(\kappa+\kappa_{i})\xi^{2}/2$ where
$\kappa_{i}=(k_{c}^{2}-k^{2}\cos^{2}\theta_{i})^{1/2}$.

The full Fourier expansion (B1) is needed in (9) in second-order
perturbation where the first-order perturbation
$\psi_{1}^{(1)}(\textbf{r})$ obtained for $z\approx0$ in (9) is
subsequently inserted in the integral with $\psi_{0}$ replaced by
$\psi_{1}^{(1)}$ to obtain $\psi_{1}^{(2)}(\textbf{r})$.

For flux calculations, we need the far field expressions for
$G(\boldsymbol\rho,z|\boldsymbol\rho',z')$ for $z>0$. Using the
two-term approximation for its z'-dependence (the last bracket in
(B3)), Eq. (15) of ref. [4] reads
\begin{equation}
G(\textbf{r}|\boldsymbol\rho',z')\cong \frac{2k_{z}}{k_{z}+i\kappa}
\frac{e^{ikr}}{r}e^{-i\textbf{k}_{\parallel}\cdot\boldsymbol\rho'}(1+\kappa
z')
\end{equation}
\pagebreak

\section{perturbation calculation up to second order}

Derivation of the results listed in section 4 is lengthy and we will
only summarize a few crucial steps, using the same symbols as in the
text.
\begin{description}
  \item[(a)] \textit{Integrations over} $\boldsymbol\rho$:\\
 By definition of $\xi(\boldsymbol\rho)$:
\begin{equation*}
  \int \xi(\boldsymbol\rho)d^{2}\boldsymbol\rho=0
\end{equation*}
\begin{equation}
\int
\xi^{2}(\boldsymbol\rho)d^{2}\boldsymbol\rho=A\left<\xi^{2}\right>=Ab^{2}
\end{equation}
By definition (1):
\begin{equation}
\int
\xi(\boldsymbol\rho)\xi(\boldsymbol\rho+\boldsymbol\delta)d^{2}\boldsymbol\rho=Af(\boldsymbol\delta)=Af(\delta)
\end{equation}
By definition (15):
\begin{equation*}
\int\int
\xi(\boldsymbol\rho)\xi(\boldsymbol\rho')e^{i\textbf{q}\cdot(\boldsymbol\rho-\boldsymbol\rho')}d^{2}\boldsymbol\rho
d^{2}\boldsymbol\rho'=A\int
f(\delta)e^{i\textbf{q}\cdot\boldsymbol\delta}d^{2}\boldsymbol\delta
\end{equation*}
\begin{equation}
=(2\pi)^{2} A F(\textbf{q})=(2\pi)^{2}A F(q)
\end{equation}
\item[(b)] \textit{UCN flux}:

Incoming flux for velocity $v_{UCN}=\frac{\hbar k}{m}$ (with
\textit{m} = neutron
  mass):
\begin{equation}
\Phi_{in}=Av_{UCN} \cos\theta_{i}
\end{equation}
Outgoing flux into solid angle $d\Omega$:
\begin{equation}
\Phi_{out}=v_{UCN}|\psi_{out}|^{2}r^{2}d\Omega
\end{equation}
For the specular beam at angle
$\Omega_{r}(\theta=\theta_{i},\varphi=0)$
\begin{equation*}
\psi_{r}\sim e^{i\textbf{k}_{r}\cdot
\textbf{r}}=e^{ikz\cos\theta_{i}}e^{ikx\sin\theta_{i}}
\end{equation*}
we can use the plane wave expansion in spherical harmonics [18] for
large distance
 $\textit{r}$:
\begin{equation}
e^{i\textbf{k}_{r}\cdot\textbf{r}}\longrightarrow
\frac{2\pi}{ik}\frac{e^{ikr}}{r}
\sum_{l=0}^{\infty}\sum_{m=-l}^{+l}Y_{l}^{m*}(\theta_{i},0)Y_{l}^{m}(\theta,\varphi)=\frac{2\pi}{ik}\frac{e^{ikr}}{r}\delta(\Omega-\Omega_{r})
\end{equation}
This allows simple integration of the interference integrals.
 \item[(c)]\textit{Loss terms}:\\
 On each step the terms containing the loss coefficient
$\eta$ are developed to first  order in $\eta$. For instance, for
$\textit{k}_{z}$ not too close to $\textit{k}_{c0}$:
\begin{equation}
\kappa=\sqrt{k_{c}^{2}-k_{z}^{2}}\cong
\kappa_{0}-\frac{i\eta}{2}\frac{k_{c0}^{2}}{\kappa_{0}}
\end{equation}
with  $k_{c}^{2}=k_{c0}^{2}(1-i\eta)$ and $\kappa_{0}=\sqrt
{k_{c0}^{2}-k_{z}^{2}}$
\end{description}
\pagebreak

\section{Integrations in $\large{\textbf{q}}$-space}
Using the transformation $d\Omega\cos\theta=\nu d\nu d\psi$ and the
relationship
\begin{equation*}
\cos^{2}\theta=1-(s_{i}-\nu)^{2}-4\nu s_{i}\sin^{2}(\psi/2),
\end{equation*}
the integrations over $\psi$ in (17)-(20) can be performed
analytically. The integrations run from 0 to $\psi_{u}$ where
\begin{equation}
\sin\frac{\psi_{u}}{2}=\left\{
\begin{array}{rl}
1; \hspace{0.3in}for\hspace{0.3in}
\nu\leq1-s\\\sqrt{\frac{1-(\nu-s_{i})^{2}}{4\nu
s_{i}}};\hspace{0.1in}for \hspace{0.1in}1+s_{i}\geq \nu \geq 1-s_{i}
\end{array} \right.
\end{equation}
Since the $\psi$-integrals needed in (18) to (20) do not appear to
be readily available from tables or websites we list two results
\begin{equation}
\int_{0}^{\psi_{u}}\cos\theta d\psi=\left\{
\begin{array}{rl}
2v_{-}^{1/2} \textbf{E}(m_{1});\hspace{0.3in} \nu\leq1-s_{i}\\2
\sigma^{1/2}\textbf{E}(m^{-1}_{1})+2v_{+}\sigma^{-1/2}
\textbf{K}(m^{-1}_{1});\hspace{0.1in}1+s_{i}\geq\nu\geq1-s_{i}
\end{array} \right.
\end{equation}
and
\begin{equation}
\int_{0}^{\psi_{u}}\frac{2\zeta^{2}\cos^{2}\theta
-1}{\sqrt{1-\zeta^{2} \cos^{2}\theta}}d\psi= \left\{
\begin{array}{rl}
2u_{+}^{-1/2}\textbf{K}(m_{2})-4u_{+}^{1/2}\textbf{E}(m_{2});\hspace{0.3in}\nu\leq1-s_{i}\\2u_{+}^{-1/2}F(\omega_{1}|m_{2})-4u_{+}^{1/2}E(\omega_{1}|m_{2})+4\zeta^{2}(-v_{-}v_{+})^{1/2};1+s_{i}\geq\nu\geq1-s_{i}
\end{array} \right.
\end{equation}
where [19]
\begin{equation}
F(\omega|m)=\int_{0}^{\omega}(1-m\sin^{2}\beta)^{-1/2}d\beta
\end{equation}
is the incomplete elliptic integral of the first kind,
\begin{equation}
E(\omega|m)=\int_{0}^{\omega}(1-m\sin^{2}\beta)^{1/2}d\beta
\end{equation}
is the incomplete integral of the second kind, and
$\textbf{K}(m)=F(\frac{\pi}{2}|m)$ and
$\textbf{E}(m)=E(\frac{\pi}{2}|m)$ are the corresponding complete
elliptic integrals of the first and second
kind, respectively. The arguments in (D2) and (D3) are \\
$\zeta=\frac{k}{k_{c0}}$; $v_{\pm}=1-(\nu\pm s_{i})^{2}$;
 $u_{\pm}=1-\zeta^{2}v_{\pm}$; $\sigma=4\nu s_{i}$; $m_{1}=\frac{\sigma}{v_{-}}$;
$m_{2}=\frac{\sigma \zeta^{2}}{u_{+}}$;
$\sin\omega_{1}=(\frac{v_{-}u_{+}}{\sigma})^{1/2}$

 Compared to
numerical double integration, the use of these readily available
functions in one of the integrals as well as tabulation of the
numerical results for use in simulations is of enormous benefit.

\pagebreak

\section{Macroscopic limit}
In the geometrical optics model, reflection is considered as an
incoherent superposition of rays reflected at each speck of a fairly
smooth surface ($\alpha<<1$) as if it were a plane inclined at the
local slope. The analysis is based on a slope distribution function
$p(\chi)$ that has to be postulated independently of the correlation
function $g(\delta)$. For instance, $p(\chi)$ does not have to be
Gaussian if $g(\delta)$ has been chosen Gaussian. If the simplest
Gaussian is chosen, as in [4], it is of form
\begin{equation}
p(\chi)=p(|\boldsymbol\chi|)=2\pi\chi
p(\boldsymbol\chi)=\frac{2\chi}{\alpha^{2}}e^{-\chi^{2}/\alpha^{2}}
\end{equation}
which is normalized and has the correct second moment $\left<
\chi^{2}\right>=\alpha^{2}$ The ambiguity is more obvious for the
$K_{0}$ model. Requiring that $p(\chi)$ should be a monotonic,
bounded function, exhibit an asymptotic \textit{K}-Bessel function
behavior and, for convenience, be amenable to analytic analysis we
can choose
\begin{equation}
p(\boldsymbol\chi)=A\frac{K_{\nu}(Z)}{Z^{\nu}}
\end{equation}
where $Z=\frac{1}{\alpha}(\chi^{2}+\beta^{2})^{1/2}$. For any $\nu
> 3/4$, values of $A$ and $\beta$ can be determined analytically [20]
such as to satisfy the normalization and second moment criteria. For
$\nu=1$: $\beta=\upsilon\alpha$ and $A=[\pi \alpha
K_{0}(\upsilon)]^{-1}$ where $\upsilon=0.1657$ is the solution to
$K_{0}(\upsilon)=2\upsilon K_{1}(\upsilon)$. For small values of
$\alpha$, the \textit{K} model, as compared with the Gauss model,
describes a surface with higher probability of large slope
($\chi>>\alpha$) at the expense of areas with a very small slope.

In the macroscopic-roughness model the probability of scattering
(local reflection) from $\Omega_{i}$ to $\Omega$, i.e. from angles
$(\theta_{i},0)$ to $(\theta,\varphi)$, at suitably oriented surface
elements is :
\begin{equation}
\frac{dP}{d\Omega}d\Omega=p(\boldsymbol\chi)d^{2}\boldsymbol\chi=p(\boldsymbol\chi)d\chi_{x}d\chi_{y}
\end{equation}
The mapping $\boldsymbol\chi\rightarrow\Omega$ involves the local
polar angle of incidence, $\theta'$, where
\begin{equation}
c'=\cos\theta'=\left(c_{+}^{2}\cos^{2}\left(\frac{\varphi}{2}\right)+c_{-}^{2}\sin^{2}\left(\frac{\varphi}{2}\right)\right)^{1/2}
\end{equation}
with $c_{\pm}=\cos[(\theta\pm\theta_{i})/2]$. We obtain
\begin{equation}
d\Omega=\frac{4c_{+}^{3}c_{-}^{3} }{c'^{2}}d^{2}\boldsymbol\chi
\end{equation}
Inserting (E5) and (E1) into (E3) gives the scattering distribution.
For the Gaussian model,
\begin{equation}
I_{sc}(\Omega_{i}\rightarrow\Omega)=\frac{dP}{d\Omega}=\frac{1}{4\pi
\alpha^{2}}\frac{c'^{2}}{c_{+}^{3}c_{-}^{3}}\exp\left[-\frac{q^{2}}{(2\alpha
k c_{+}c_{-})^{2}}\right]
\end{equation}
where the in-plane momentum transfer \textit{q} is given in (11).

Eq. (E6) is consistent with the limiting case analyzed in ref. [4]
(Eq. (26) of [4]) and it closely resembles the micro-roughness
result (14) for $\textit{I}_{(11)}$ (for $\eta=0$). It exhibits a
similar width of scattering distribution around the regular
reflection angle $(\theta_{i},0)$ but no exactly specular intensity.
Local angle $(\theta')$ dependent loss could be incorporated in the
form $\mu_{0}(\theta')$ as for a plane surface, but the result is
not identical to the micro-roughness result in the limit $\alpha <<
1$.

Expression (E6) and similar expressions for non-Gaussian models are
symmetric in $\Omega_{i}$ and $\Omega$, thus do not satisfy the
detailed balance requirement discussed in Section 6, and indeed give
rise to loss of isotropy in computer simulations. Moreover, (E6) is
catastrophically inadequate for glancing-angle incidence
($\theta_{i}\rightarrow\pi/2$) due to shadowing and multiple
reflection within a rough surface as discussed in Section 6.
Although this angular range is small, it makes a large contribution
to the mean value for isotropic incident UCN flux and cannot be
neglected.


\begin{thebibliography}{9}
\bibitem{12}
Using a Distorted Wave Born Approximation, the expression
$\exp(-4b^{2}kk'\cos\theta_{i}cos\theta_{i}')$ was obtained in [1]
and [5], where the primed quantities refer to the quantities inside
the refractive medium. This modification is clearly inapplicable
within the total reflection range, where $k'\cos\theta_{i}'$ is
imaginary (approaching zero at the edge).
\end{thebibliography}
\end{document}